\documentclass[superscriptaddress,nofootinbib,amsmath,amssymb,prc,showpacs,twocolumn,floatfix]{revtex4-2}

\usepackage{graphicx}
\usepackage{dcolumn}
\usepackage{bm}
\usepackage{xcolor}
\usepackage{ulem}

\usepackage[colorlinks]{hyperref}
\hypersetup{linkcolor=red,citecolor=blue,urlcolor=blue}

\usepackage{float,subfig,slashed,hyperref}

\begin{document}

\title{Thermal twin stars within a hybrid equation of state based on a nonlocal chiral quark model compatible with modern astrophysical observations}

\author{J.P. Carlomagno}
\affiliation{CONICET, Godoy Cruz 2290, Buenos Aires, Argentina}
\affiliation{IFLP, UNLP, CONICET, Facultad de Ciencias Exactas, Diagonal 113 entre 63 y 64, La Plata 1900, Argentina}
\author{G.A. Contrera}
\affiliation{CONICET, Godoy Cruz 2290, Buenos Aires, Argentina}
\affiliation{IFLP, UNLP, CONICET, Facultad de Ciencias Exactas, Diagonal 113 entre 63 y 64, La Plata 1900, Argentina}
\author{A.G. Grunfeld}
\affiliation{CONICET, Godoy Cruz 2290, Buenos Aires, Argentina}
\affiliation{Departamento de F\'\i sica, Comisi\'on Nacional de Energ\'{\i}a At\'omica, Av. Libertador 8250, (1429) Buenos Aires, Argentina}
\author{D. Blaschke}
\affiliation{Institute for Theoretical Physics, University of Wroclaw, Max Born Pl. 9, 50-204, Wroclaw, Poland}
\affiliation{Helmholtz-Zentrum Dresden-Rossendorf (HZDR), Bautzner Landstrasse 400, 01328 Dresden, Germany}
\affiliation{Center for Advanced Systems Understanding (CASUS), Untermarkt 20, 02826 G\"orlitz, Germany}

\begin{abstract}
We investigate the extension to finite temperatures and neutrino chemical potentials of a recently developed nonlocal chiral quark model approach to the
equation of state of neutron star matter.
We consider two light quark flavors and current-current interactions in the
scalar-pseudoscalar, vector, and diquark pairing channels, where the nonlocality of the currents is taken into account by a Gaussian form factor that depends on the spatial components of the 4-momentum.
Within this framework, we analyze order parameters, critical temperatures, phase diagrams, equations of state, and mass-radius relations for different
temperatures and neutrino chemical potentials.
For parameters of the model that are constrained by recent multi-messenger observations of neutron stars, we find that the mass-radius diagram for isothermal hybrid star sequences exhibits the thermal twin phenomenon for temperatures above 30 MeV.
\end{abstract}

\maketitle

\section{Introduction}

The exploration of the QCD phase diagram remains a focal point of research, drawing considerable attention due to its implications for understanding the diverse phases of strongly interacting matter with temperature and baryon chemical potential~\cite{Fukushima:2010bq}. In extreme conditions, such as those encountered in the early universe or within neutron stars~\cite{Schwarz:2003du, Page:2006ud, Lattimer:2015nhk, Blaschke:2018mqw},
nuclear matter undergoes transitions across various phases, encompassing the quark-gluon plasma, hadronic matter, and color superconducting phases
\cite{Blaschke:2018mqw,Baym:2017whm}.
Nevertheless, when QCD transitions to the non-perturbative regime at low energies, effective models emerge as indispensable tools for elucidating relevant phenomena in this domain~\cite{Vogl:1991qt, Klevansky:1992qe, Hatsuda:1994pi, Buballa:2003qv}. In regimes characterized by low temperatures and high densities or baryonic chemical potential, it is imperative that effective models proficiently encapsulate and predict phenomena within compact stellar environments. Conversely, near-zero baryonic chemical potential, effective models must align with the insights provided by lattice QCD (LQCD)~\cite{Karsch:2001vs, Ding:2015ona}. This highlights how complex QCD is, and we need flexible theories to understand it at various energy and density levels.
The growing wealth of data from different cosmic messengers highlights the need to thoroughly test effective models. This ensures accurate predictions and descriptions of various phenomena. 
In particular, it is important to note that nonlocal quark effective models, specifically with covariant form factors, naturally result in the inverse magnetic catalysis effect under a constant external magnetic field and finite temperature conditions \cite{Pagura:2016pwr, Ali:2020jsy}.
Additionally, it's interesting to explore how temperature and neutrino trapping play a role, especially when describing objects formed after compact star mergers. This also includes considering thermal twins \cite{Hempel:2015vlg}, which can be crucial in simulating supernovae for massive stars. Integrating these factors helps us build more comprehensive models for a better understanding of astrophysical intricacies.
In our prior research \cite{Contrera:2022tqh}, we conducted a thorough analysis of a nonlocal chiral quark model, incorporating color superconductivity and vector repulsive interactions at zero temperature. The primary aim was to leverage the resulting equation of state (EOS) for quark matter (QM) in conjunction with a hadronic EOS to comprehensively characterize the properties of cold, deleptonized neutron stars (NS) within a hybrid framework. We determined the optimal parameters of the QM model that satisfy modern observational constraints, including maximum mass, radii, and tidal deformability \cite{LIGOScientific:2018cki, Miller:2021qha, Hebeler:2013nza, Ayriyan:2021prr}.

In this ongoing investigation, our central objective is to enhance the model's versatility at high densities by extending its applicability to finite temperature conditions.
Then, the developed EOS can become part of the repository CompOSE \cite{Antonopoulou:2022yot} which comprises EOS for simulations of astrophysical objects like supernovae, neutron stars and their mergers.

To maintain consistency with constraints pertinent to the cold neutron star (NS) scenario, we employ the identical parameters from the quark matter (QM) model detailed in Ref. \cite{Contrera:2022tqh}. By incorporating temperature, we consider the presence of trapped neutrinos in compact star matter, examining their collective impact on the maximum mass and radii of the compact object configurations.

This paper is organized as follows: In Section \ref{sect:3DFFqm}, we introduce the quark model, extending its application to finite temperature. Subsequently, in Section \ref{sect:AA}, we present the results of our hybrid model for astrophysical applications. Finally, Section \ref{SC} provides a summary of our findings and conclusions.

\section{Instantaneous nonlocal quark model at finite temperature and density, including neutrino trapping}
\label{sect:3DFFqm}

We investigate the properties of quark matter (QM) in the frame of a nonlocal chiral quark model. This model takes into account interactions involving scalar and vector quark-antiquark pairs, as well as anti-triplet scalar diquark interactions. The effective Euclidean Lagrangian density for two light flavors is expressed as ~\cite{Contrera:2022tqh}
\begin{eqnarray}
\mathcal{L} &=&  \bar \psi (x) \left(- i \rlap/\partial + m_c
\right) \psi (x) - \frac{G_S}{2} j^f_S(x) j^f_S(x)
\nonumber \\
&-&  \frac{G_D}{2} \left[j^a_D(x)\right]^\dagger j^a_D(x)
{+} \frac{G_V}{2} j_V^{\mu}(x)\, j_V^{{\mu}}(x) \ ,
\label{action}
\end{eqnarray}
In this context, $m_c$ denotes the current quark mass, which is assumed to be the same for both $u$ and $d$ quarks. The currents $j_{S,D}(x)$ are defined using nonlocal operators based on a separable approximation of the effective one-gluon exchange (OGE) model within the framework of Quantum Chromodynamics (QCD).~\cite{Blaschke:2007ri, GomezDumm:2005hy}.
The currents read
\begin{eqnarray}
j^f_S (x) &=& \int d^4 z \  g(z) \ \bar \psi(x+\frac{z}{2}) \ \Gamma_f\, \psi(x-\frac{z}{2})\,,
\nonumber \\
j^a_D (x) &=&  \int d^4 z \ g(z)\ \bar \psi_C(x+\frac{z}{2}) \ i \gamma_5 \tau_2 \lambda_a \ \psi(x-\frac{z}{2})\,,
\nonumber \\
{j^\mu_V (x)} &=& {\bar \psi(x)~ i\gamma^\mu\ \psi(x)}\,,
\label{cuOGE}
\end{eqnarray}
where we defined $\psi_C(x) = \gamma_2\gamma_4 \,\bar \psi^T(x)$ and $\Gamma_f=(\openone,i\gamma_5\vec\tau)$, while $\vec \tau$ and $\lambda_a$, with $a=2,5,7$, stand for Pauli and Gell-Mann matrices acting on flavor and color spaces, respectively.
The functions $g(z)$ in Eqs.~(\ref{cuOGE}) represent nonlocal ``instantaneous'' form factors (3D-FF) characterizing the effective quark interaction, which relies on spatial momentum components.

It is important to mention that the vector current in Eq.~(\ref{cuOGE}) is assumed to be local, and the reason for this assumption will be provided later.
For the extension to 2+1 flavors, see Refs.~\cite{Blaschke:2005uj,Ruester:2005jc}.

In the mean-field (MF) approximation, the only nonvanishing MF values in the scalar and vector sectors correspond to isospin zero fields, specifically $\bar\sigma$ and $\bar\omega$, respectively.
Additionally, within the diquark sector, due to color symmetry, one can perform color space rotations to set $\bar\Delta_5$ and $\bar\Delta_7$ to zero while maintaining $\bar\Delta_2 = \bar\Delta$.

When distinct chemical potentials, denoted as $\mu_{fc}$ for each flavor and color, are introduced, it may initially appear as though six unique quark chemical potentials emerge. These correspond to the quark flavors $u$ and $d$, as well as the quark colors $r$, $g$, and $b$.
However, all $\mu_{fc}$ can be expressed in terms of three independent parameters: the baryonic chemical potential $\mu_B$ (calculated as $\mu_u + 2\mu_d$), a quark electric chemical potential $\mu_{Q_q}$ (defined as $\mu_u - \mu_d$), and a color chemical potential $\mu_8$.
The corresponding relations read
\begin{eqnarray}
\mu_{ur} &=& \mu_{ug} = \frac{\mu_B}{3} + \frac23 \mu_{Q_q} + \frac13
\mu_8 \nonumber \\
\mu_{dr} &=& \mu_{dg} = \frac{\mu_B}{3} - \frac13 \mu_{Q_q} + \frac13
\mu_8 \nonumber \\
\mu_{ub} &=& \frac{\mu_B}{3} + \frac23 \mu_{Q_q} -
\frac23 \mu_8 \nonumber \\
\mu_{db} &=& \frac{\mu_B}{3} - \frac13 \mu_{Q_q} -
\frac23 \mu_8 .
\label{chemical}
\end{eqnarray}
Note that the particular choice of $\bar\Delta_2 = \bar\Delta$ in the space color introduces an additional $r-g$ symmetry.

When taking into account the vector meson mean field $\bar{\omega}$, originated from the term involving $\gamma_0$ in the local vector current, the chemical potentials experience a shift denoted as $\tilde{\mu}_{fc} = \mu_{fc} - \bar{\omega}$~\cite{Buballa:2003qv}.
This choice of local interactions in the vector current was made to prevent any momentum dependence in the chemical potentials, which are now adjusted due to the presence of the vector mean field.
In addition, following Ref.~\cite{Blaschke:2003yn}, it is convenient to define
\begin{equation}
\tilde{\mu}_c = \frac{\tilde{\mu}_{uc} + \tilde{\mu}_{dc}}{2}
\end{equation}
and
\begin{equation}
\delta\tilde{\mu}_c = \frac{\tilde{\mu}_{uc} - \tilde{\mu}_{dc}}{2}.
\end{equation}

Thus, the corresponding mean field grand canonical thermodynamic potential per unit volume can be written as
\begin{eqnarray}
\Omega^{MFA}  = \frac{ \bar
\sigma^2 }{2 G_S} + \frac{ {\bar \Delta}^2}{2 G_D} - \frac{\bar
\omega^2}{2 G_V}\ - 2 \int \frac{d^3 \vec{p}}{(2\pi)^3} \;
\xi(\vec{p}) ,
\label{mfaqmtp}
\end{eqnarray}
with
\begin{eqnarray}
\xi(\vec{p})  & &=  \sum_{c,\kappa,\lambda=\pm}  \; \left\{ \epsilon _c^\kappa/2 \; + \; T \; \ln \left[
1 \; + \; e^{- \tfrac{\epsilon _c^\kappa + \lambda \ \delta\tilde\mu_c}{T}}\right] \right\} .
\label{integrando}
\end{eqnarray}
where $c$ denotes quark color and the $\pm$ for $\kappa$ and $\lambda$ indicate the need to consider two terms for each index, one for each sign.
Considering the red-green symmetry introduced earlier, we have defined

\begin{equation}
\epsilon _c^{\pm} = \bar{E}_c^{\pm} \sqrt{1 + {\left[{g(\vec{p}) \bar \Delta (1-\delta_{bc})}/{\bar{E}_c^{\pm}}\right]^2}}
\end{equation}
where
\begin{equation}
\bar{E}_c^{\pm} = E \pm \tilde{\mu}_c.
\end{equation}
The dispersion relation is given by
\begin{equation}
E^{2} = \vec{p}~^2 + M^2(\vec{p}).
\end{equation}
Here, the momentum-dependent quark mass function is
\begin{equation}
M(\vec{p}) = m_c + g(\vec{p}) \bar\sigma.
\end{equation}

The mean field values $\bar \sigma$ and $\bar \Delta$ are determined by solving a system of coupled gap equations, complemented by a constraint equation for $\bar \omega$. This set of equations collectively characterizes the self-consistent behavior of the system. Additionally, the constraint equation for $\bar \omega$ ensures that the vector meson mean field remains consistent with the other field values, contributing to the overall stability and equilibrium of the system under investigation. The set of equations is:
\begin{eqnarray}
\frac{ \partial \Omega^{MFA}}{\partial \bar \sigma} = 0 \ , \ \ \
\frac{\partial \Omega^{MFA}}{\partial \bar \Delta} = 0 \ , \ \ \
\frac{\partial \Omega^{MFA}}{\partial \bar \omega} = 0 \
\label{gapeq}
\end{eqnarray}
explicitly shown in  Eqs.~(\ref{gap_sig})-(\ref{gap_ovec}), and using the regularization prescription of Eq.~(\ref{eq:omreg}).

As we aim to describe quark matter behavior in the core of neutron stars, we need to consider the presence of leptons. In this study, we consider exclusively electrons and electron neutrinos as the leptonic components. By treating leptons as a free relativistic Fermi gas, the total pressure of both quark matter and leptons can be expressed as:
\begin{equation}
P = P_{q} + P_{lep} \ ,
\label{potential}
\end{equation}
with $P_q = - \Omega^{MFA}_{reg}$ (see Appendix~\ref{sect:details_QM}),  and where $P_{lep}$ reads
\begin{eqnarray}
P_{lep} &=& 2  \; T\; \sum_{\lambda =\pm}
\int \frac{d^3\vec{p}}{(2 \pi)^3} \ln \left[
1 \; + \; e^{- \tfrac{\epsilon_e + \lambda \; \mu_e}{T}}\right]  \nonumber \\
&+&
\left(\frac{\mu_{\bar \nu_e}^4}{24\pi^2} + \frac{\mu_{\bar \nu_e}^2\ T^2}{12} + \frac{7\pi^2\ T^4}{360} \right) \ ,
\label{lepfree}
\end{eqnarray}
with $\epsilon_e = \sqrt{\vec{p}^2 + m_e^2} $. %

In addition, it is necessary to take into account that quark matter has to be in $\beta$ equilibrium with electrons and muons through the $\beta$-decay reactions
\begin{equation}
d\to u+e+\bar\nu_e\ ,~~~  u+e\to d+\nu_e .
\end{equation}
Thus, we have an additional relation between fermion chemical potentials, namely,
\begin{equation}
\mu_{Q_q} = \mu_{uc} - \mu_{dc} = - \mu_e + \mu_{\nu_e}
\label{munu}
\end{equation}
for $c=r,g,b$.

Ensuring electric and color charge neutrality within the system is a crucial requirement in the core of neutron stars. In this context, two chemical potentials, namely $\mu_e$ and $\mu_8$, become constrained by the conditions that electric charge and color charge number densities must be zero. As we will introduce later, we will consider that $\mu_{\nu_e}$ is a function of the temperature. These conditions are expressed as follows:
\begin{eqnarray}
n_{Q_{tot}} &=& n_{Q_q}- n_e \nonumber \\
& =& \sum_{c=r,g,b} \left(\frac23 \ n_{uc} - \frac13 \ n_{dc} \right)
- n_e \ = \ 0 \ \ , \nonumber \\
n_8 & = & \frac{1}{\sqrt3} \sum_{f=u,d}
\left(n_{fr}+n_{fg}-2n_{fb} \right) \ = \ 0 \ ,
\label{dens}
\end{eqnarray}
where the expressions for the different number densities can be found in Appendix~\ref{sect:details_QM}.

From the thermodynamic potential, we can easily derive several other important quantities.
In particular, we define the quark and lepton densities as follows:
\begin{eqnarray}
n_{fc}  &=&  - \frac{\partial \Omega^{MFA}_{reg}}{\partial \mu_{fc}} \ , \nonumber \\
n_e &=& - \frac{\partial \Omega^{MFA}_{reg}}{\partial \mu_{e}} \ ,\nonumber \\
n_{\nu_e} &=& \frac{\mu_{\bar \nu_e}^3}{8\pi^2} + \frac{\mu_{\bar \nu_e}\ T^2}{6} \ .
\label{densities}
\end{eqnarray}
The quark chiral condensate and chiral susceptibility $\chi$ are given by
\begin{eqnarray}
\langle \bar \psi \psi \rangle  =  \frac{ \partial \Omega^{MFA}_{reg}}{\partial m_c} \ , ~~
\chi = - \frac{ \partial \langle \bar \psi \psi \rangle} {\partial m_c}\ .
\end{eqnarray}

In summary, within the context of quark matter in neutron stars, it is possible to determine the values of $\bar \Delta$, $\bar \sigma$, $\bar \omega$, $\mu_e$, and $\mu_8$, for each combination of temperature ($T$) and baryonic chemical potential ($\mu_B$). This is achieved through the solution of Eqs.~(\ref{gapeq}), accompanied by the supplementary equations (\ref{munu}) and (\ref{dens}).
This comprehensive approach enables us to establish the equation of state (EOS) for quark matter within the specific thermodynamic regime.

To comprehensively define the nonlocal NJL model in question, it is essential to establish specific parameters and the instantaneous form factor $g(\vec{p})$ at $T=0$ and $\mu_B=0$. These parameters and form factors are vital for describing how quarks interact in the $q\bar{q}$ and $qq$ channels.
As in Ref.~\cite{Contrera:2022tqh}, we consider a Gaussian
form factor in momentum space,
\begin{eqnarray}
\label{ff}
g(\vec{p}) &=& \exp[-\vec{p}\,^2/\Lambda^2]\ . \nonumber
\end{eqnarray}
In this study, we adopted the identical set of input parameters as presented in Ref.~\cite{Contrera:2022tqh}, which includes $m_c = 2.3$~MeV and $G_S=9.9$~GeV$^{-2}$.
We introduce the ratios of coupling constants as follows: $\eta_D = G_D/G_S$ and $\eta_V = G_V/G_S$. We will specifically focus on the ratios explored in Ref.~\cite{Contrera:2022tqh}, which were constrained through the analysis of observational multi-messenger data.

Now, we can begin to analyze the features of the phase transitions in the $T$ - $\mu_B$ plane for the nonlocal chiral quark model introduced above. In this section, we will simplify our analysis by neglecting the influence of neutrino trapping. The consideration of neutrinos and their impact, as a function of the temperature, will be incorporated in the subsequent section when we investigate astrophysical applications. This separation allows us to focus on the specific aspects of the system at hand before introducing the broader astrophysical context.

\begin{figure}[h]
    \centering
    \includegraphics[width=0.48\textwidth]{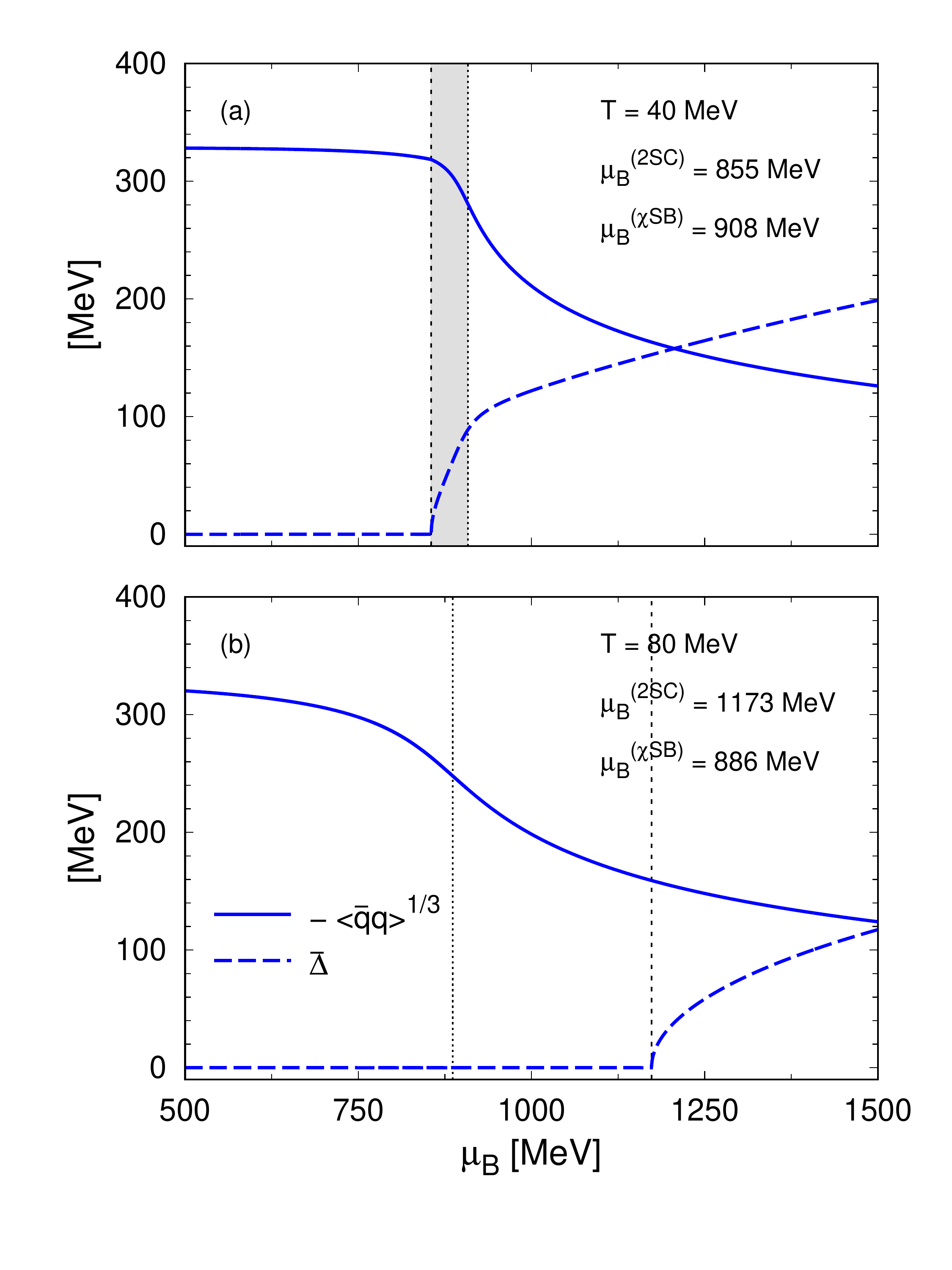}
    \caption{Order parameters chiral condensate (solid line) and diquark pairing gap (dashed line) as functions of $\mu_B$ for temperatures $T=40$~MeV and $T=80$~MeV in panels (a) and (b), respectively, when $\eta_D=1.1$, $\eta_V=0.5$ and $\mu_{\nu_e}=0$. }
    \label{figOPa}
\end{figure}

To illustrate the behavior of the order parameters and phase transitions, it is convenient to start by showing the behavior of the order parameters for representative values of $\eta_D$ and $\eta_V$ as a function of the baryonic chemical potential, considering two different fixed temperatures.
Initially, we will consider only electrons as leptons, that is, without neutrinos trapped in the system.
In Fig.~\ref{figOPa}, we quote the quark condensate in solid lines and the diquark MF value in dashed lines as functions of $\mu_B$, for $\eta_D=1.1$ and $\eta_V=0.5$.
The critical chemical potentials $\mu_B^c$ are denoted with thin black vertical lines.

\begin{figure}[h]
    \centering
    \includegraphics[width=0.48\textwidth]{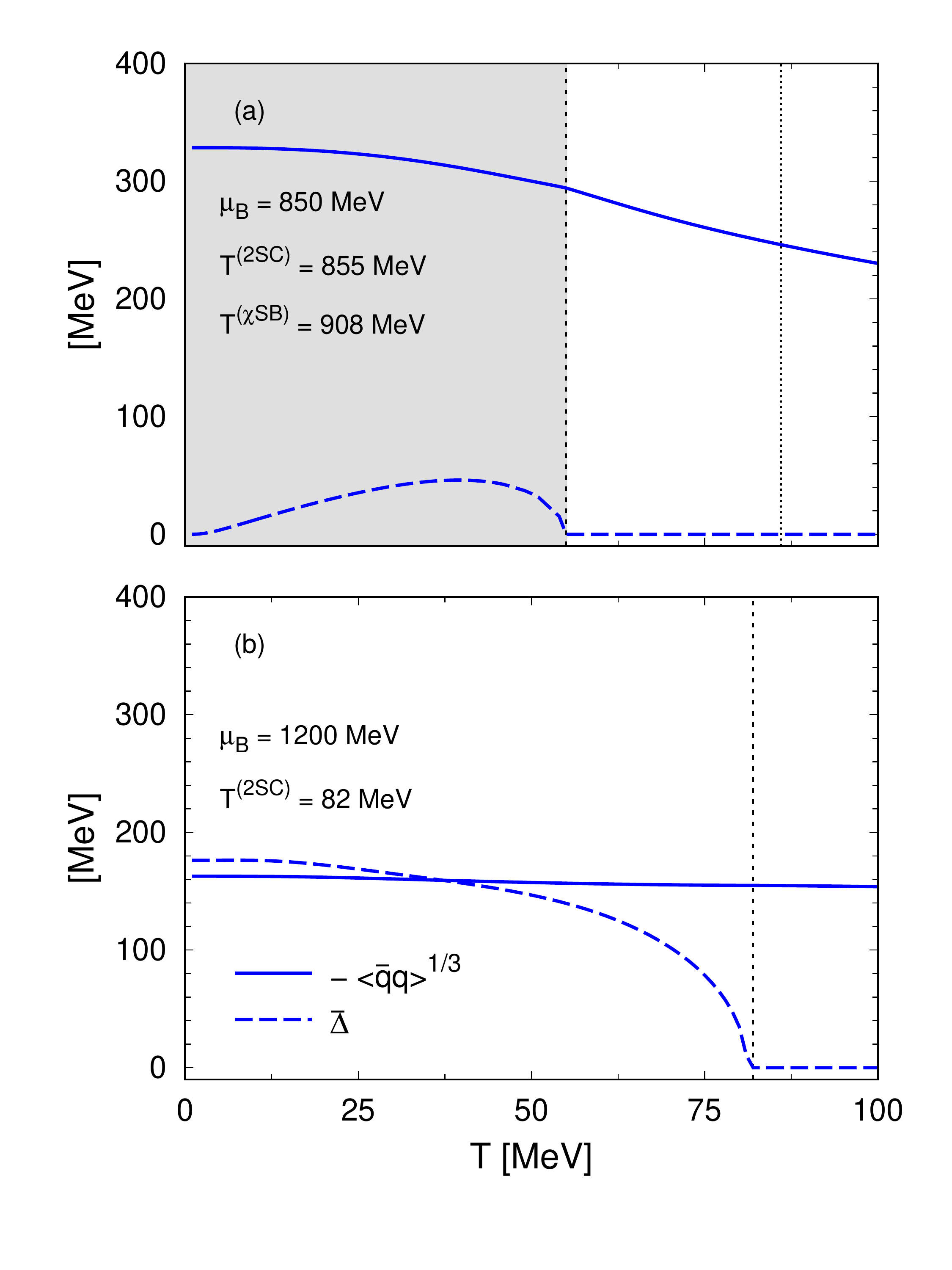}
    \caption{Order parameters chiral condensate (solid line) and diquark pairing gap (dashed line) as functions of temperature $T$ for $\mu_B=850$~MeV and $\mu_B=1200$~MeV in panels (a) and (b), respectively, with $\eta_D=1.1$, $\eta_V=0.5$ and $\mu_{\nu_e}=0$. }
    \label{figOPb}
\end{figure}

In Fig.~\ref{figOPb} we show the behavior of the quark condensate in solid lines and the diquark MF value in dashed lines as functions of $T$ for two given values of $\mu_B$.

In general, it can be seen that the mean field value of the diquark field vanishes at $\mu_B^{2SC}$ ($T^{2SC}$), denoting a second-order phase transition.
On the other hand, at $\mu_B^{\chi SB}$ ($T^{\chi SB}$), one finds the peak of the chiral susceptibility, indicating a crossover phase transition to a region where the chiral symmetry is partially restored.
Therefore, one can define a region, denoted by the gray band in both figures, where a 2SC phase takes place with a finite and small value of the diquark gap coexisting with the chiral symmetry-breaking phase.

\begin{figure}[h]
    \centering
    \includegraphics[width=0.45\textwidth]{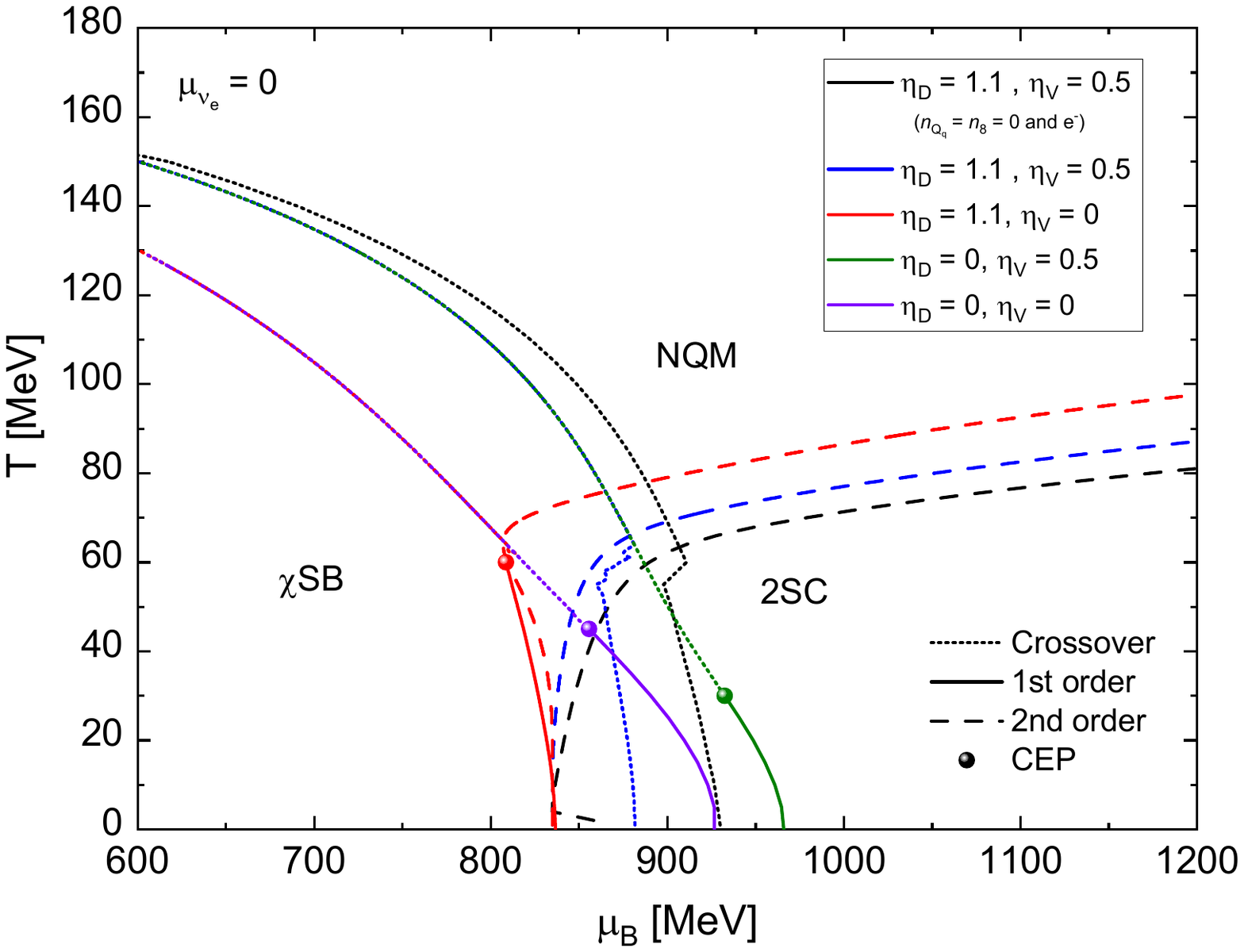}
    \caption{
    Structure of the phase diagram at low temperatures and finite $\mu_B$ for different parametrizations of the model:
    full model with leptons (black line), version without charge or color neutrality (blue lines), and the simplest version, a model without diquarks or vector interactions (violet lines). }
    \label{figvsPD}
\end{figure}

Considering the phase diagrams under different model approximations is valuable. It not only facilitates comparisons with other studies but also aids in understanding the impact of each interaction term considered on the phase transition curves.
In Fig.~\ref{figvsPD}, we present the phase diagrams corresponding to different limiting conditions of the model, depending on the choice of the coupling constant ratios.

The phase diagram can be sketched by analyzing the numerical results obtained for the relevant order parameters.
In general, one can find regions in which the chiral symmetry is either broken ($\chi$SB) or approximately restored, and phases in which the system remains either in an asymptotically free phase (NQM) or in a two-flavor superconducting phase (2SC).
The first-order and crossover boundaries are depicted by solid and dotted lines, respectively, while second-order phase transitions are indicated by dashed curves. Solid dots mark critical end point (CEP) locations, between the crossover and first-order phase transitions.
Each set of lines of the same color represents the phase diagram for a specific approximation of the model.
Namely, the full model is shown with black lines, while the blue ones correspond to a system without color and charge neutrality or leptons. In red (green) lines we plot the phase transition curves for a model with $\eta_V=0$ ( $\eta_D=0$).
Finally, the purple curves represent a model with $\eta_V = \eta_D = 0$.
It is evident that diquark interactions promote chiral symmetry restoration, whereas vector repulsion appears to delay it. Furthermore, as the diquark coupling constant ratio increases, the 2SC region is more robust, as expected.

\begin{figure}[h]
    \centering
    \includegraphics[width=0.44\textwidth]{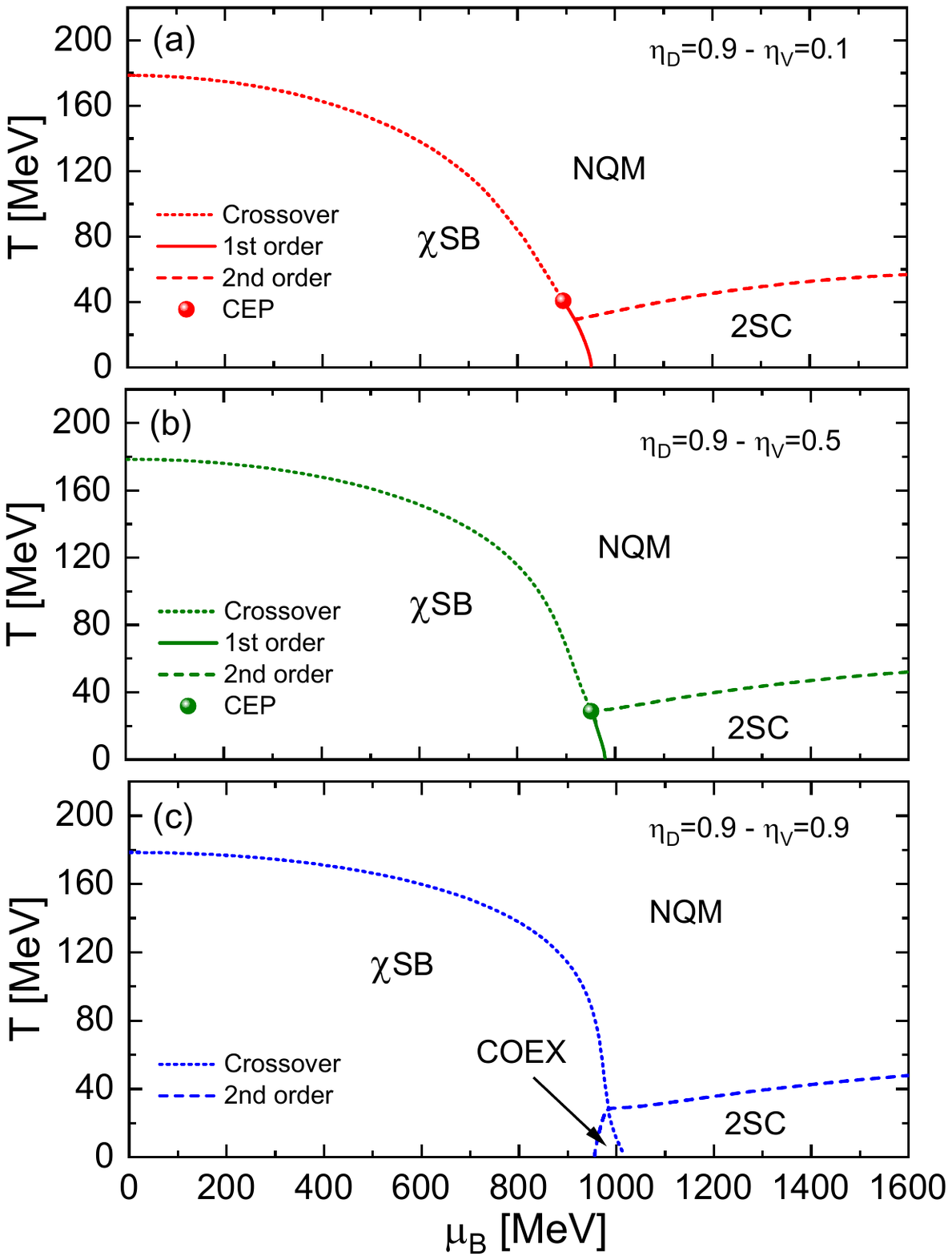}
    \caption{Phase diagrams for $\eta_D=0.9$ and (a) $\eta_V=0.1$, (b) $\eta_V=0.5$ and (c) $\eta_V=0.9$ with $\mu_{\nu_e}=0$}
    \label{figPD1}
\end{figure}

\begin{figure}[h]
    \centering
    \includegraphics[width=0.44\textwidth]{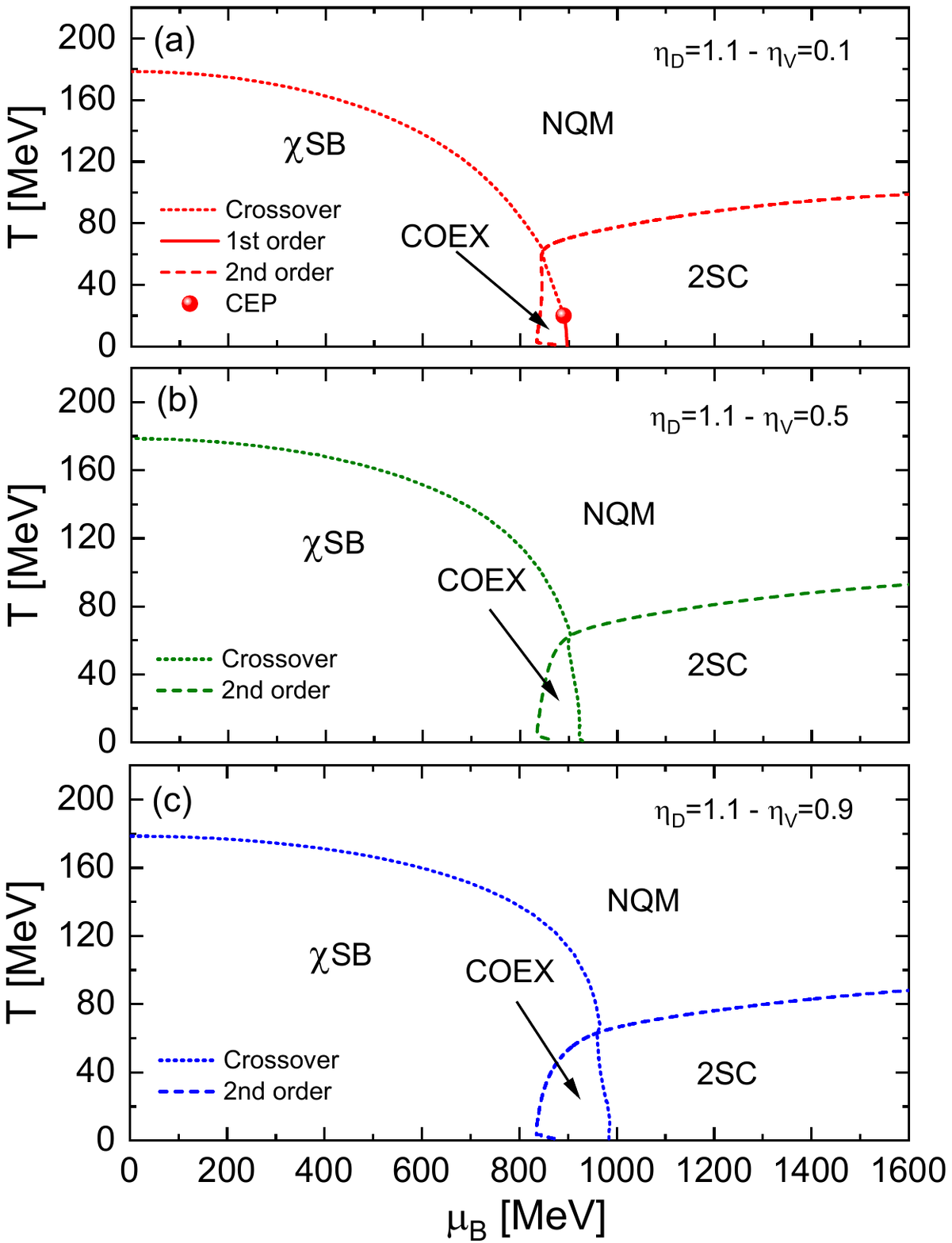}
    \caption{Phase diagrams for $\eta_D=1.1$ and (a) $\eta_V=0.1$, (b) $\eta_V=0.5$ and (c) $\eta_V=0.9$ with $\mu_{\nu_e}=0$.}
    \label{figPD2}
\end{figure}

Let us now analyze the structure of the phase diagram at the mean-field level for several representative cases in the parameter space of $\eta_D$ and $\eta_V$, as considered in Ref.~\cite{Contrera:2022tqh}. Specifically, we fix $\eta_D$ to 0.9 in Fig.~\ref{figPD1} and to 1.1 in Fig.~\ref{figPD2}. In these figures, the upper, middle, and lower panels correspond to $\eta_V$ values of 0.1, 0.5, and 0.9, respectively.
At relatively high temperatures, the critical chemical potentials are characterized by the positions of the peaks in the chiral susceptibility, marking the region where the transition occurs as a smooth crossover (denoted by dotted lines in the figures).
Conversely, at lower temperatures, the chiral condensate exhibits a discontinuity, indicating a first-order phase transition
(solid lines in the figures).
Traversing the first-order phase transition curve reveals an ascent in the critical temperature from zero to a critical end point (CEP) temperature, while the critical chemical potential decreases as $T_c$ increases. Beyond this CEP, the chiral restoration phase transition proceeds smoothly as a crossover.

With increasing vector coupling, the CEP is moved to lower temperatures and is absent above a critical vector coupling, as was found already in \cite{Sasaki:2006ws}.
As previous investigations of the QCD phase diagram with color superconductivity within chiral quark models at the mean-field level have revealed (see, e.g., Refs.~\cite{Blaschke:2005uj, Ruester:2005jc, Abuki:2005ms}), the chiral symmetry breaking and color superconducting phases expel each other when vector and diquark couplings are not too strong.
Then, the first-order phase transition of partial chiral symmetry restoration for $T_c<T_{\rm CEP}$ entails a first-order transition for the onset of color superconductivity with increasing $\mu_B$.  
Should vector and/or diquark coupling be sufficiently strong, then a coexistence phase of chiral symmetry breaking and color superconductivity can exist. This has been observed and discussed in the three-flavor case in the work \cite{Hatsuda:2006ps} as a result of the mixing between diquark pairing condensates and chiral condensates by the Fierz-transformed 't Hooft determinant interaction, i.e. as a consequence of the $U_A(1)$ anomaly of QCD.
This coexistence phase is also related to the possibility of a 
BEC-BCS crossover in low-temperature quark matter \cite{Blaschke:2008uf, Zablocki:2009ds, Abuki:2010jq, Zablocki:2010zz}.
Note that a crossover nature of the deconfinement transition at low temperatures has also been suggested based on arguments for a quark-hadron continuity \cite{Schafer:1998ef, Wetterich:1999vd}.

As mentioned above, there is an intermediate region of coexistence where the chiral symmetry remains broken with a nonvanishing diquark mean-field value. The size of the coexistence region becomes larger with increasing values of $\eta_V$ and $\eta_D$. 
In contrast, elevated $\eta_V$ values lead to a reduction in the coordinates of the CEP.

Moreover, in the case of extreme $\eta_D$ values and very low temperatures, there is a sharp increase in the critical chemical potential for diquark condensation. This phenomenon, resulting from particularly large $\eta_D$ values, has been previously observed within the framework of these effective models~\cite{GomezDumm:2005hy}.

\section{Astrophysical applications}
\label{sect:AA}

The primary objective of this study is to assess the applicability of the proposed QM model in the context of astrophysics, in particular to the study of effects of the quark-hadron phase transition in (proto-)neutron stars as well as simulations of supernova explosions and neutron star mergers.
In this context, the appearance of thermal twins \cite{Hempel:2015vlg} signals a softening of the EOS which may inhibit the possibility of a second shock that was obtained as an explosion mechanism for the DD2F-SF model class of hybrid EOSs \cite{Fischer:2017lag, Jakobus:2022ucs} and rather lead to a failed supernova.
The possibility that thermal twins might serve as an indicator for the (non-)explodability property of a class of hybrid EOSs warrants a detailed study of this phenomenon.
To achieve this goal, we adopt a two-phase model framework for the construction of isothermal first-order phase transitions from hadronic matter to QM under the constraints met in the interiors of compact stellar objects.

\subsection{QM EOS}
\label{QMEoS}

In Ref. \cite{Contrera:2022tqh} we considered a $\mu_B$-dependent bag pressure given by the equation
\begin{equation}
B(\mu_B) = B_0 \, f_< (\mu_B)
\end{equation}
with
\begin{equation}
f_< (\mu_B) = \frac{1}{2}\left[1 - \mathrm{tanh} \left( \frac{\mu_B - \mu_<}{\Gamma_<} \right)\right],
\end{equation}
that was first introduced in \cite{Alvarez-Castillo:2018pve} in order to mimic the string-flip QM EOS of Ref.~\cite{Kaltenborn:2017hus} and later applied also for other aspects of hybrid neutron star phenomenology \cite{Blaschke:2020qqj, Shahrbaf:2020uau, Shahrbaf:2022upc}.

Here we have set $\mu_< = 895$ MeV, $\Gamma_< = 180$ MeV and $B_0 = 35$ MeV/fm$^3$, being the optimal values to reproduce the astrophysical observables.

Also, it is important to remark that a $\mu_B$-dependent bag pressure will affect the value of $n_B={\partial P}/{\partial \mu_B}$, producing a noticeable effect on the Gibbs free energy and therefore also on the energy density, both of them defined below.
The energy density can be written as
\begin{equation}
\varepsilon = -P^* + T\;s + G
\end{equation}
where $s = -\partial \Omega/\partial T$.
Here, $G$ is Gibbs free energy, which depends on conserved charges
\begin{eqnarray}
G &=& \sum_{\alpha} \mu_{\alpha} \; n_{\alpha}  \nonumber \\
&=& \mu_B \; n^*_B + \mu_l \; n_l + \mu_Q \; n_Q + \mu_8 \; n_8.
\label{Gibbs}
\end{eqnarray}
where $n^*_B = n_B - \frac{\partial B(\mu)}{\partial \mu_B}$ and $P^* = P - B(\mu)$ and
\begin{equation}
\frac{\partial B(\mu)}{\partial \mu_B} =
- \frac{0.5~B_0}{\cosh(\frac{\mu_B-\mu_<}{\Gamma_<})^2 \,  \Gamma_<}.
\end{equation}

By imposing electric charge and color charge neutrality, the last two terms of Eq.~(\ref{Gibbs}) are zero; then, after reordering terms, $G$ can be written as
\begin{equation}
G = n^*_B \; \mu_B  + (n_{\nu_{e}} + n_e)\; \mu_{\nu_{e}}
\label{eq:Gibbs_energy}
\end{equation}
where $n_B = (1/3)(n_u + n_d)$ with $n_f = \sum_c n_{fc}$ and the chemical potentials $\mu_{fc}$ are defined in Eqs.~(\ref{chemical}).

\subsection{Hadronic phase}
\label{sect:hadrmodel}

The interactions between baryons in the hadronic phase of nuclear matter are modeled using the density-dependent relativistic mean-field (DDRMF) theory. This theory is based on the exchange of scalar ($\sigma$), vector ($\omega$), and isovector ($\rho$) mesons. The Lagrangian density of this model is a function of the meson and baryon fields, given by
\begin{eqnarray}
  \mathcal{L} &=& \sum_{B}\bar{\psi}_B \bigl[\gamma_\mu [i\partial^\mu
  - g_{\omega B}(n)
    \omega^\mu
    - g_{\rho B}(n) {\boldsymbol{\tau}} \cdot {\boldsymbol{\rho}}^\mu]
    \nonumber\\
     & -&
   \label{eq:Blag}
     [m_B - g_{\sigma B}(n)\sigma]
    \bigr] \psi_B + \frac{1}{2} (\partial_\mu \sigma\partial^\mu
  \sigma  - m_\sigma^2 \sigma^2)\\
   &-&
  \frac{1}{4}\omega_{\mu\nu} \omega^{\mu\nu} + \frac{1}{2}m_\omega^2\omega_\mu \omega^\mu + \frac{1}{2}m_\rho^2
  {\boldsymbol{\rho\,}}_\mu \cdot {\boldsymbol{\rho\,}}^\mu \nonumber\\
  &-&
  \frac{1}{4}{\boldsymbol{\rho\,}}_{\mu\nu} \cdot {\boldsymbol{\rho\,}}^{\mu\nu} \, ,\nonumber
\end{eqnarray}
where $n = \sum_B n_B$ the total baryon number density, while $g_{\sigma B}(n)$, $g_{\omega B}(n)$ and $g_{\rho B}(n)$ are density-dependent meson-baryon coupling constants, whose functional form is usually given by \cite{Typel:1999yq, Typel:2018cap}
\begin{equation}
g_{\tau B}(n) = g_{\tau B}(n_0)\,\, a_{\tau}\,
\frac{1+b_{\tau}(\frac{n}{n_0}+d_{\tau})^{2}}{1+c_{\tau}(\frac{n}{n_0}+d_{\tau})^{2}}\, ,
\end{equation}
for $\tau=\sigma,\omega$ and
\begin{equation}
g_{\rho B}(n) = g_{\rho B}(n_0)\,\mathrm{exp}\left[\,-a_{\rho} \left(\frac{n}{n_0} -  1\right)\,\right]\, ,
\end{equation}
where the parameters $a_{\tau}$, $b_{\tau}$, $c_{\tau}$, $d_{\tau}$ and $a_{\rho}$ are fixed by the binding energies, charge and diffraction radii, spin-orbit splittings, and the neutron skin thickness of finite nuclei \cite{Malfatti:2019tpg, Spinella:2020gkw}.

The meson mean-field equations following from Eq.\ (\ref{eq:Blag}) are given by
\begin{eqnarray}
m_{\sigma}^2 \bar{\sigma} &=& \sum_{B} g_{\sigma B}(n) n_B^s \, , \nonumber\\ m_{\omega}^2 \bar{\omega} &=& \sum_{B}
g_{\omega B}(n) n_{B}\, , \\ m_{\rho}^2\bar{\rho} &=& \sum_{B}g_{\rho
  B}(n)I_{3B} n_{B} \, , \nonumber
\end{eqnarray}
where $I_{3B}$ is the 3-component of isospin for each baryon, while $n_{B}^s$ and $n_{B}$ are the scalar and particle number densities for each baryon $B$, which are given by
\begin{eqnarray}
n_{B}^s&=& \gamma_B \int \frac{d^3p}{(2 \pi)^3} \left[f_{B-}(p) -
  f_{B+}(p)\right] \frac{m_B^*}{E_B^*}, \\ n_{B}&=& \gamma_B \int
\frac{d^3p}{(2 \pi)^3} \left[f_{B-}(p) - f_{B+}(p)\right] \, .
\end{eqnarray}
Here $f_{B\mp}$ denotes the Fermi-Dirac distribution function and $E^*_B$ stands for the effective baryon energy given by
\begin{equation}
f_{B\mp}(p)=\frac{1}{\exp\left[\frac{E_B^*(p) \mp \mu_B^*}{T}\right] +
  1},\,\,\,\, E_B^*(p)=\sqrt{p^2 + m_B^{*2}}\, ,\nonumber
\end{equation}
$\gamma_B=2J_B+1$ is the spin degeneration factor, $m_B^*= m_B - g_{\sigma B}(n)\bar{\sigma}$ is the effective baryon mass and $\mu_B^*$ is the effective chemical potential, given by
\begin{equation}
\mu_B^* = \mu_B - g_{\omega B}(n) \bar{\omega} - g_{\rho B}(n)
\bar{\rho} I_{3B} - \widetilde{R} \, ,
\end{equation}
where $\widetilde{R}$ is the rearrangement term given by
\begin{eqnarray}
\widetilde{R} =\sum_B&&\left( \frac{\partial g_{\omega B}(n)}{\partial
  n} n_B \bar{\omega} + \frac{\partial g_{\rho B}(n)}{\partial n}
I_{3B} n_B \bar{\rho} \right. \\ &-& \left.\frac{\partial g_{\sigma
    B}(n)}{\partial n} n_B^s \bar{\sigma}\right) \, , \nonumber
\label{rear}
\end{eqnarray}
which is important for achieving thermodynamical consistency \cite{Hofmann:2001}.  This term also contributes to the baryonic pressure,
\begin{eqnarray}
P_B &=& \sum_B \frac{\gamma_B}{3} \int \frac{d^3p}{(2 \pi)^3}
\frac{p^2}{E_B^*} [f_{B-}(p) + f_{B+}(p)] \nonumber\\ &-& \frac{1}{2}
m_{\sigma}^2 \bar{\sigma}^2 + \frac{1}{2} m_{\omega}^2 \bar{\omega}^2
+ \frac{1}{2} m_{\rho}^2 \bar{\rho}^2 + n \widetilde{R}.
\label{HM:pressure}
\end{eqnarray}

Note that in this work, neglecting the effects of strangeness, $B$ runs only for nucleons, i.e., $B={N, P}$.

Lepton particles can be included in the hadronic matter as free Fermi gases in the relativistic mean-field (RMF) theory. The pressure contribution of these particles is given by
\begin{equation}
P_L = \sum_L \frac{\gamma_L}{3} \int \frac{d^3p}{(2 \pi)^3}
\frac{p^2}{E_L} [f_{L-}(p) + f_{L+}(p)] \, ,
\label{eq:leptons}
\end{equation}
where $f_{L\mp}(p)$ are the corresponding Fermi-Dirac distribution functions for leptons and antileptons, respectively.
The degeneracy factor of the spin-1/2 leptons is $\gamma_L=2$.

The sum over $L$ in Eq.\ (\ref{eq:leptons}) usually runs over $e^-$ with mass $m_e$ and, when corresponding, massless electron neutrinos, $\nu_e$.

The energy density, $\varepsilon$, is determined by the Gibbs relation:
\begin{equation}
 \varepsilon = - P + T S + \sum_j \mu_j \, n_j \, ,
 \label{eq:EoS}
\end{equation}
where $P=P_B+P_L$, $S = \frac{\partial P}{\partial T}$ and $n_j = \frac{\partial P}{\partial \mu_j}$ ($j$ stands for all the particles of this phase, including leptons).

As previously shown in the $T=0$ calculation by Contrera {\it{et al}}.~\cite{Contrera:2022tqh}, our study incorporates a neutron star crust, utilizing the Baym-Pethick-Sutherland (BPS) model \cite{Baym:1971pw} to comprehensively characterize the hadronic equation of state (EOS) at densities below the nuclear saturation density.

\subsection{Exploring the hybrid EOS and astrophysical observables}

To obtain the mass-radius relations of the proto-neutron stars we use a two-phase description to account for the transition from nuclear matter to quark matter (QM).
For QM we use the nonlocal NJL model presented in Sec.~\ref{sect:3DFFqm} and studied at $T=0$ in Ref.~\cite{Contrera:2022tqh}, which includes a density-dependent bag pressure and whose free parameters have been chosen to better reproduce modern astrophysical constraints.
On the other hand, to describe nuclear matter at finite temperature, we use the DD2 density-dependent model parametrization described in Sec.~\ref{sect:hadrmodel}.

The phase transition between nuclear and quark matter is described by a Maxwell construction, where it is required that the pressure and Gibbs free energy per baryon of the two phases coincide at the phase transition. Note that, at $T=0$ without neutrino trapping, the Gibbs free energy per baryon becomes the baryon chemical potential, as shown in Eq. (\ref{eq:Gibbs_energy}).
Outside the phase transition, the phase with higher pressure and lower Gibbs free energy per baryon has to be chosen as the physical one.

To assess and compare the hybrid equation of state (EOS) with astrophysical observations, it is necessary to solve the Tolman-Oppenheimer-Volkoff (TOV) equations for a static, non-rotating, spherically symmetric star.
Specifically, to compute the internal energy density distribution of compact stars and thus derive the mass-radius relation we utilize the TOV equations for a static and spherical star in the framework of general relativity:
\begin{align}
    \frac{dP(r)}{dr} &= \frac{G(\varepsilon(r))+P(r))(M(r)+4\pi r^3 P(r))}{r(r-2GM(r))},\\
    \frac{dM(r)}{dr} &= 4\pi r^2 \varepsilon(r)\,,
    \label{eq:TOV}
\end{align}
with $P(r=R)=0$ and $P(r=0)=P_c$ as boundary conditions for a star with mass $M$ and radius $R$.

In a neutron star at finite temperature, neutrinos are trapped in the stellar core. In this work, we neglect muons and muon neutrinos due to their low fractions and negligible impact on global stellar properties~\cite{Prakash:1996xs, Chiapparini:1995xd, Steiner:2000bi, Shao:2011nu, Chen:2013}.
Therefore, we will consider only electrons and the corresponding (anti)neutrinos as leptons for both hadronic and quark phases.
We consider that the neutrino chemical potential is a linear function of the temperature, where three different scenarios can be identified (following Ref.~\cite{Lugones:2021tee}): (i) NS with extremely low temperatures and no trapped neutrinos, (ii) proto-NS (PNS) exhibiting low temperatures and a significant quantity of trapped neutrinos, and (iii) post-merger object (PMO) could reach high temperatures, with the significant neutrino trapping amount.

Considering the parameters associated with the quark model, we establish the values for the ratios of the coupling constants as $\eta_D=1.1$ and $\eta_V=0.5$.
In our analysis, we delineate the phase diagram for QM, considering the assumption that trapped neutrinos demonstrate a linear dependency on temperature. This ansatz is precisely described by $\mu_{\nu_e} = 2T$. We considered the impact of trapped neutrinos by including their chemical potential in Eq. (\ref{munu}). It's essential to use Eq. (\ref{munu}) together with the chemical potentials provided in Eq. (\ref{chemical}).

\begin{figure}[h]
    \centering
    \includegraphics[width=0.48\textwidth]{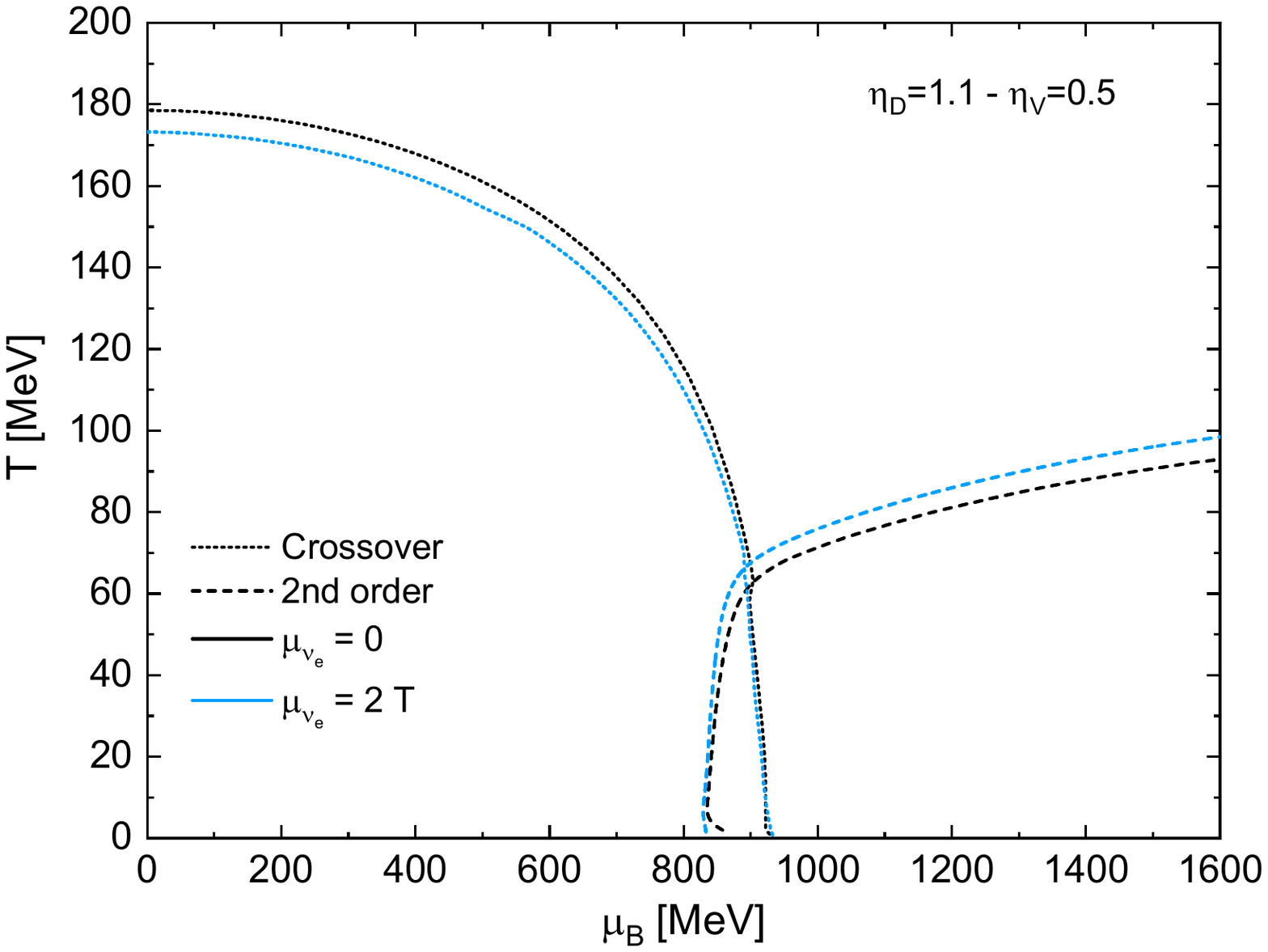}
    \caption{Phase diagram for quark matter model with $\eta_V = 0.5$ and $\eta_D = 1.1$ with and without neutrino trapping.}
    \label{figPDmunu}
\end{figure}

In Fig.~\ref{figPDmunu}, we compared the phase diagrams with and without neutrinos, shown as blue and black lines, respectively.
As expected, at very low temperatures, both diagrams overlap, owing to our ansatz for the temperature dependence of the neutrino chemical potential which implies that at $T=0$ neutrino trapping effects are absent.
At higher temperatures, with the inclusion of neutrinos in the model, we observed a slight reduction of the chiral critical temperature, accompanied by a marginally expanded 2SC phase, see also Refs. \cite{Ruester:2005ib,Sandin:2007zr}.

\begin{figure}[h]
    \centering
    \includegraphics[width=0.48\textwidth]{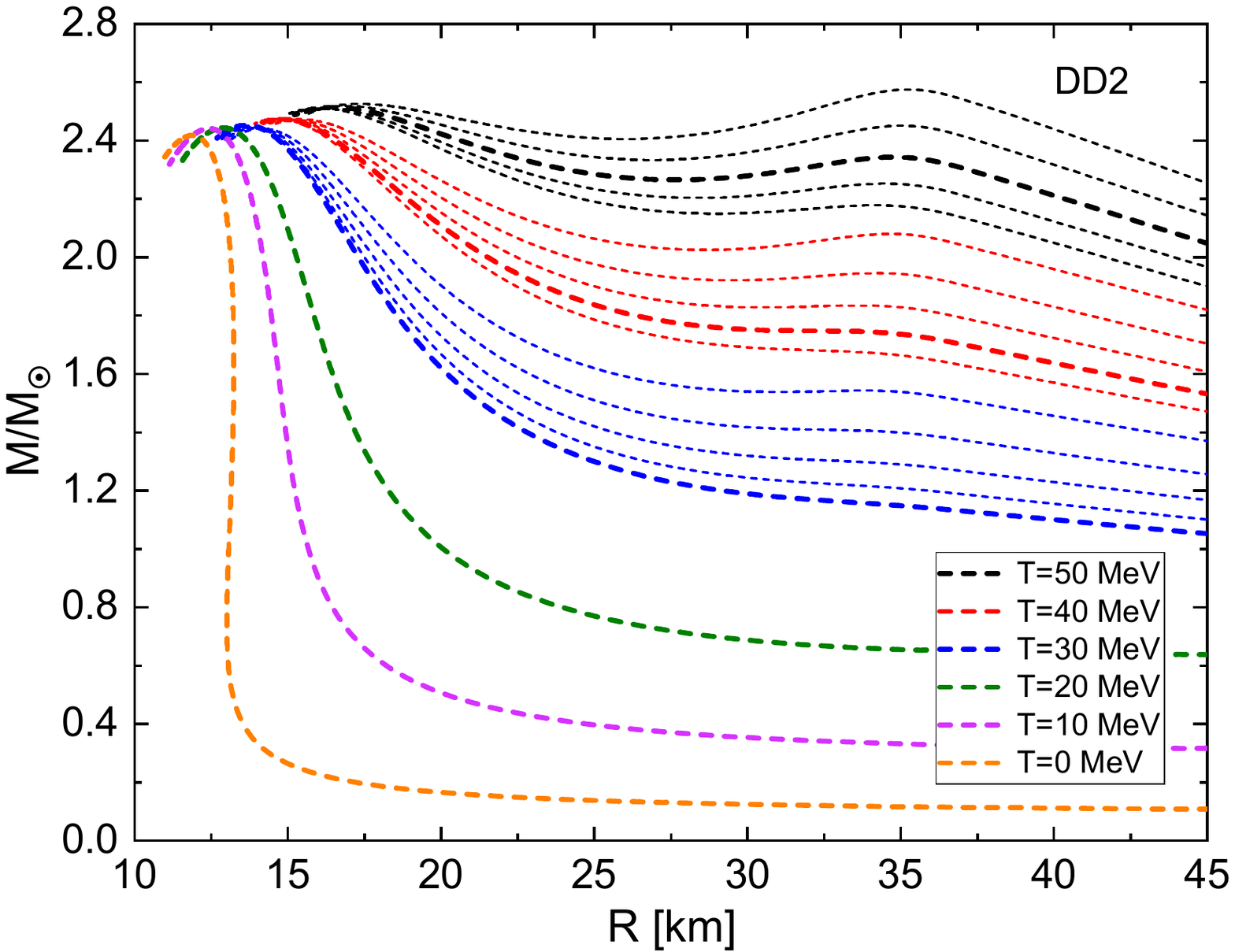}
    \caption{ Mass-Radius relations for pure hadronic matter EOS (with a BPS crust). Different values of $T$ and the corresponding $\mu_{\nu_{e}} = 2 T$ are highlighted with thicker lines. In addition to the existing data points, for $T=$ 30, 40, and 50 MeV we added thinner lines corresponding to additional fixed values of $\mu_{\nu_{e}} = 60, 80, 100, 120, 140$ MeV, from bottom to top.}
    \label{fig7}
\end{figure}

In Fig.~\ref{fig7}, the mass-radius plot for the hadronic matter is presented, using the DD2 EOS with a BPS crust. Three distinct relevant temperatures are considered alongside various selections of neutrino chemical potentials. It can be seen that the maximum mass of hadronic compact stars increases with both temperature and neutrino chemical potential. An expansion in the radii accompanies this trend. As the temperature rises, possibly indicating a post-merger state, a second maximum is noticeable, characterized by a larger radius. This observation may suggest the emergence of an additional family of expanded (twin) neutron stars at high temperatures.

\begin{figure}[h]
    \centering
    \includegraphics[width=0.48\textwidth]{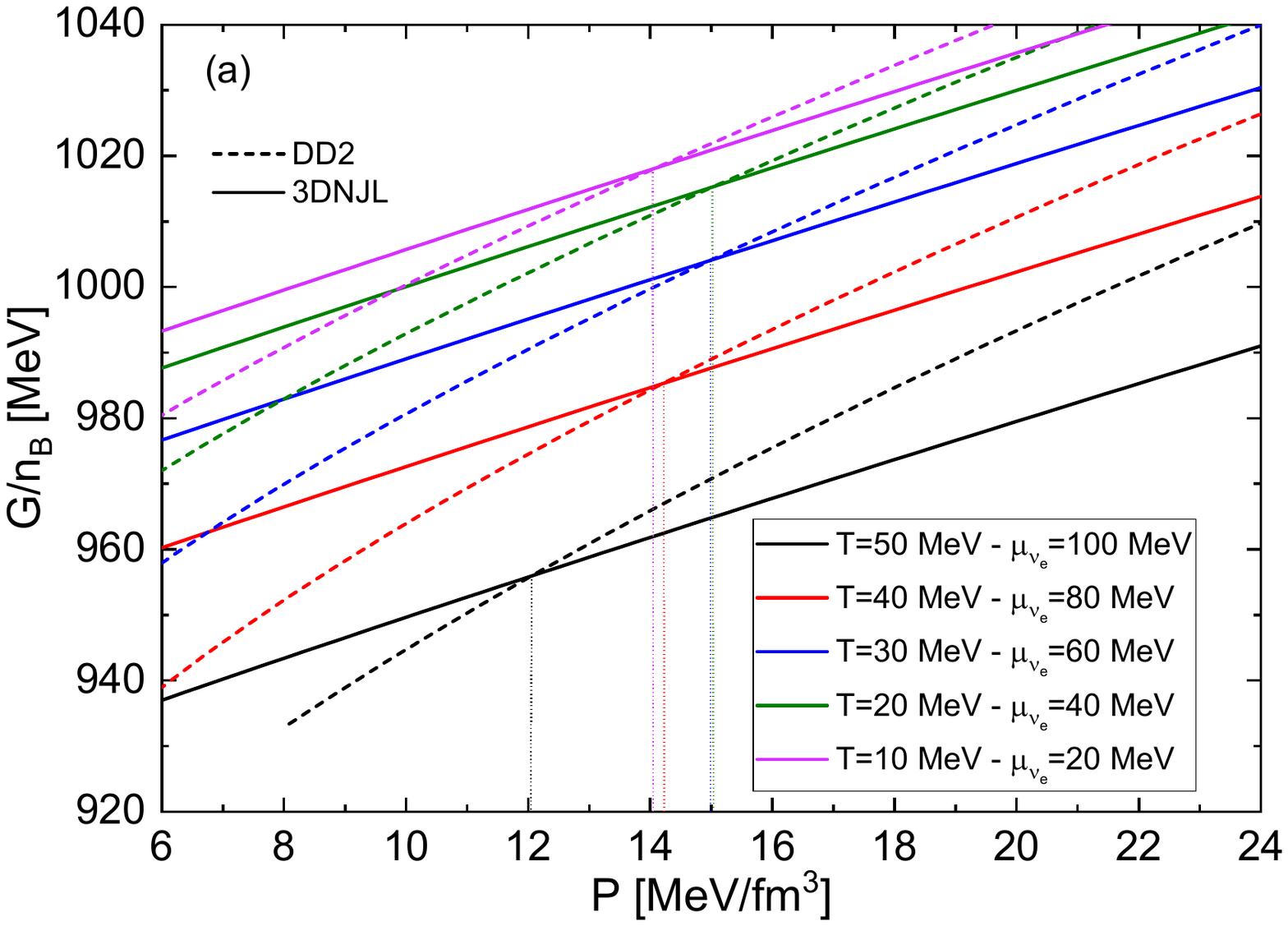}
    \includegraphics[width=0.48\textwidth]{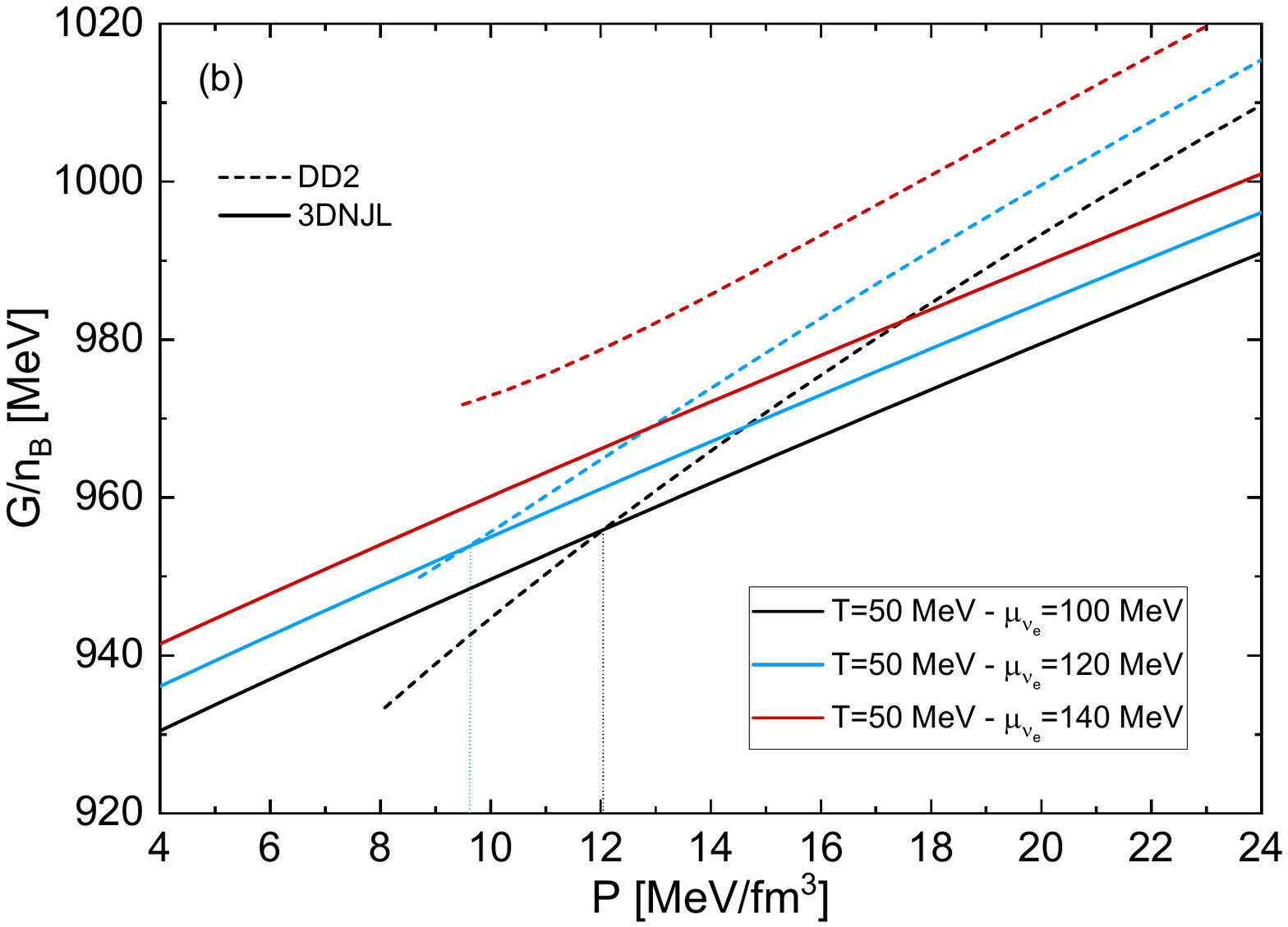}
    \caption{ (a) Hadron-Quark phase transition points for finite temperature EOS considering the relation $\mu_{\nu_e}/T = 2$  and (b) $\mu_{\nu_e}/T = \alpha$, with $\alpha = 2.4$ and $2.8$.}
    \label{fig8}
\end{figure}

Now, we will analyze the hybrid EOS configurations, composed of a DD2 hadronic phase + BPS crust and a nonlocal 3DFF model for the QM phase with input parameters $\eta_V = 0.5$ and $\eta_D = 1.1$. First, in Fig.~\ref{fig8} (a), we present the Gibbs free energy per baryon as a function of pressure to illustrate the transitions from the hadronic to quark matter phase. We examined a range of $T$ and $\mu_{\nu_{e}}$ values, which are linked, as previously mentioned, by $\mu_{\nu_e} = 2T$.
Subsequently, in Fig.~\ref{fig8} (b), we depict the Gibbs free energy per baryon concerning pressure while considering an alternative linear relation for $\mu_{\nu_e} = \alpha T$. As an illustrative example, we consider values of $\alpha$ equal to 2.4 and 2.8. It is apparent that with an increase in $\alpha$, the crossing between the hadronic and QM EOS disappears, resulting in the absence of hybrid configurations.

\begin{figure}[h]
    \centering    \includegraphics[width=0.48\textwidth]{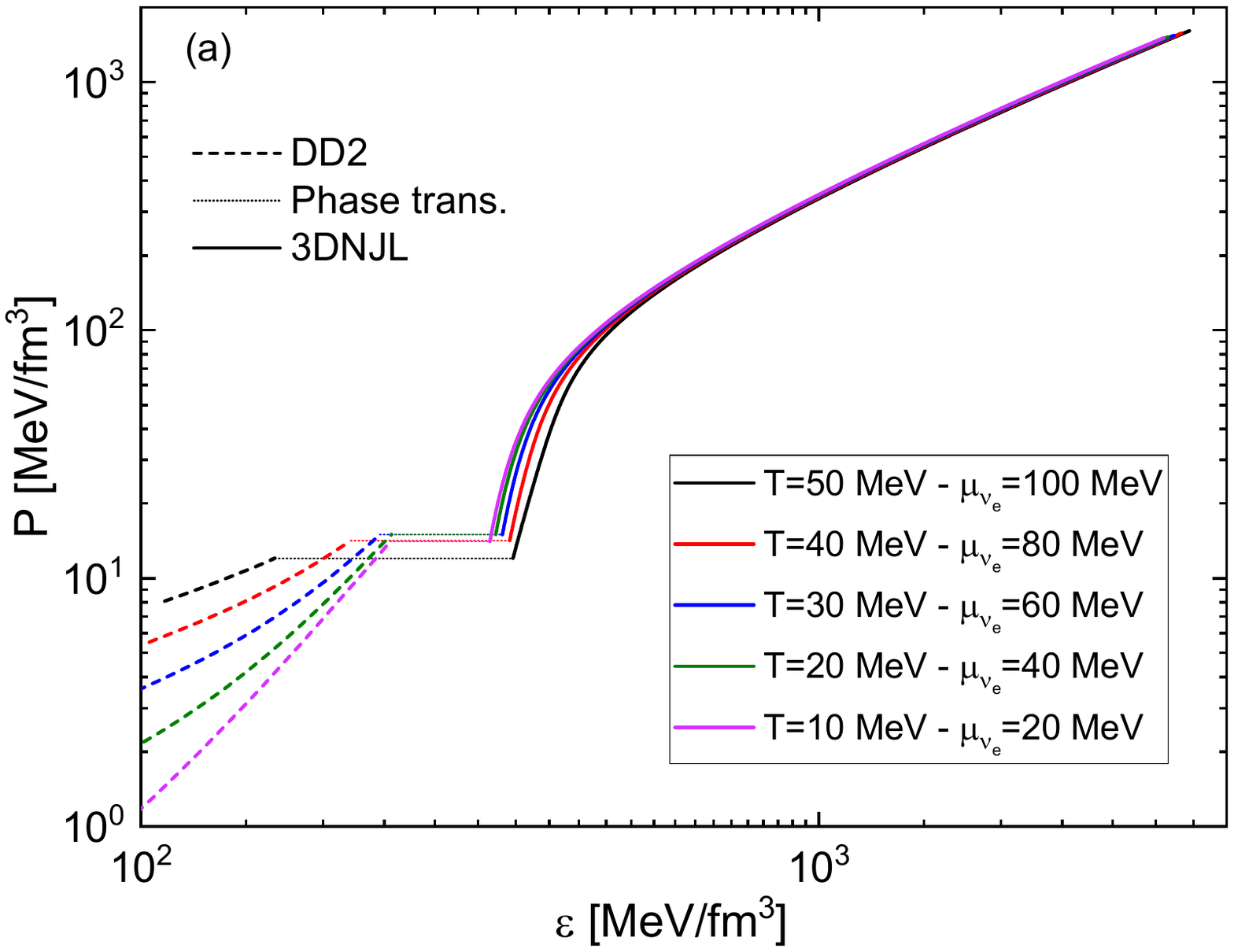}
    \includegraphics[width=0.48\textwidth]{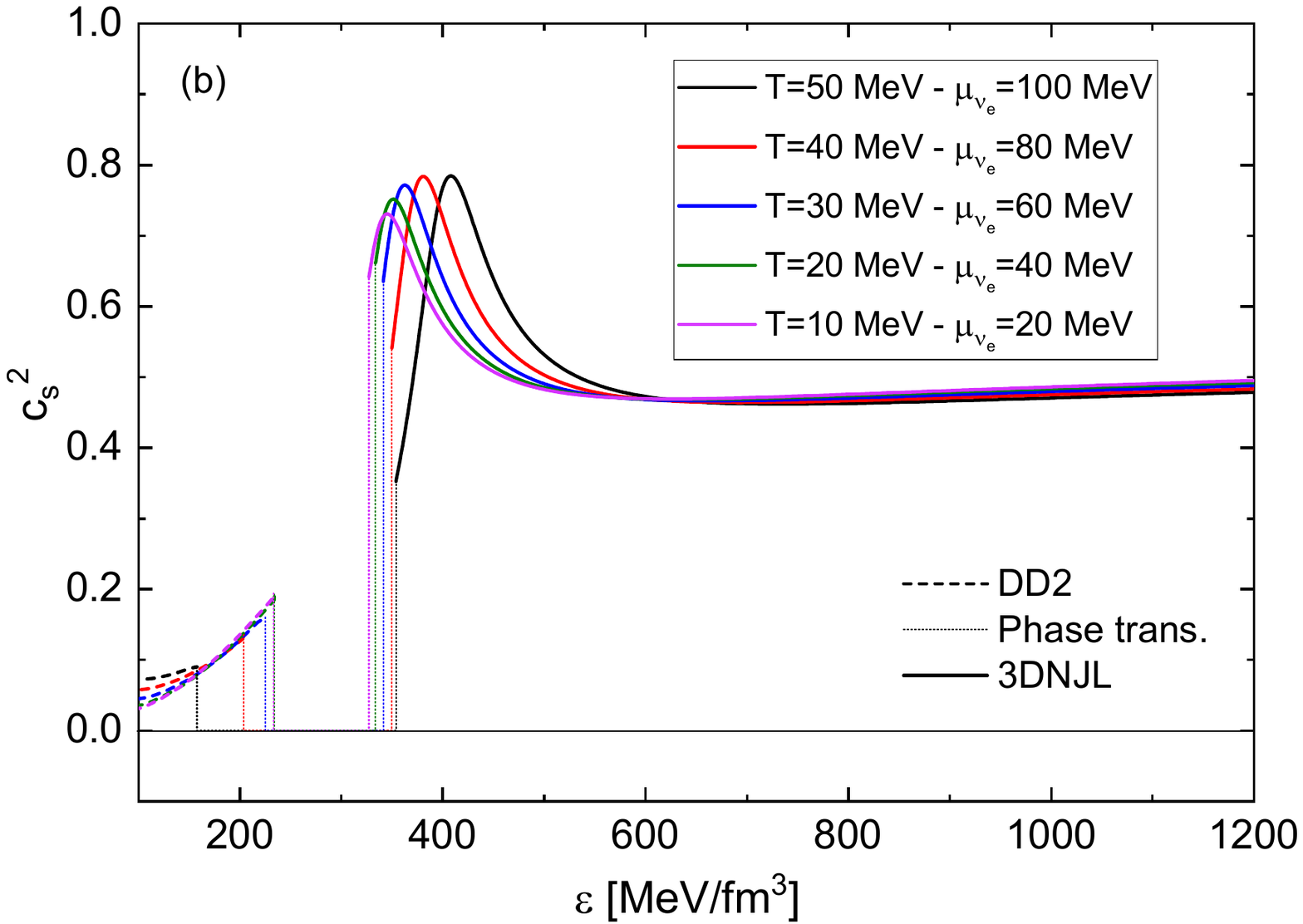}
    \caption{(a) Hybrid EOS constructed from the phase transitions shown in Fig.~\ref{fig8} (a). (b) Squared speed of sound as a function of the energy density, for several T (and the related $\mu_{\nu_e}$). }
    \label{fig9}
\end{figure}

In Fig.~\ref{fig9} (a), we show the hybrid EOS resulting from a Maxwell construction of the phase transition as shown in Fig.~\ref{fig8} (a). We observe that as the temperature increases, the jump in energy density at the transition from hadronic to quark matter is enlarged. As expected, at high densities all the QM EOSs tend to have the same linear relation between $P$ and $\varepsilon$.
In Fig.~\ref{fig9} (b), we illustrate the behavior of the squared speed of sound as a function of energy density for the hybrid EOS. It is evident that the peaks consistently remain within the causal regime.

We note that the strong first-order phase transition occurs for energy densities $\varepsilon \approx 0.2 - 0.35$ GeV/fm$^3$, just below the position of the peaks at $\sim 0.4$ GeV/fm$^3$.
This finding is supported by the recent result within an independent hybrid neutron star description based on a relativistic confining density functional model of color superconducting quark matter with an asymptotic approach to the conformal limit \cite{Ivanytskyi:2022bjc}.
Interestingly, the range of energy densities where the crossover transition is seen in lattice QCD simulations at finite temperature lies in the same range, see Fig. 1 of Ref.~\cite{Alvarez-Castillo:2014dva}.
Indications for a dip in the squared speed of sound just below the peak position have been found also by model-agnostic Bayesian analyses of the mass and radius constraints from modern astrophysical observations of neutron stars, see
\cite{Marczenko:2022jhl,Brandes:2023hma,Annala:2023cwx}.

\begin{figure}[h]
    \centering    \includegraphics[width=0.48\textwidth]{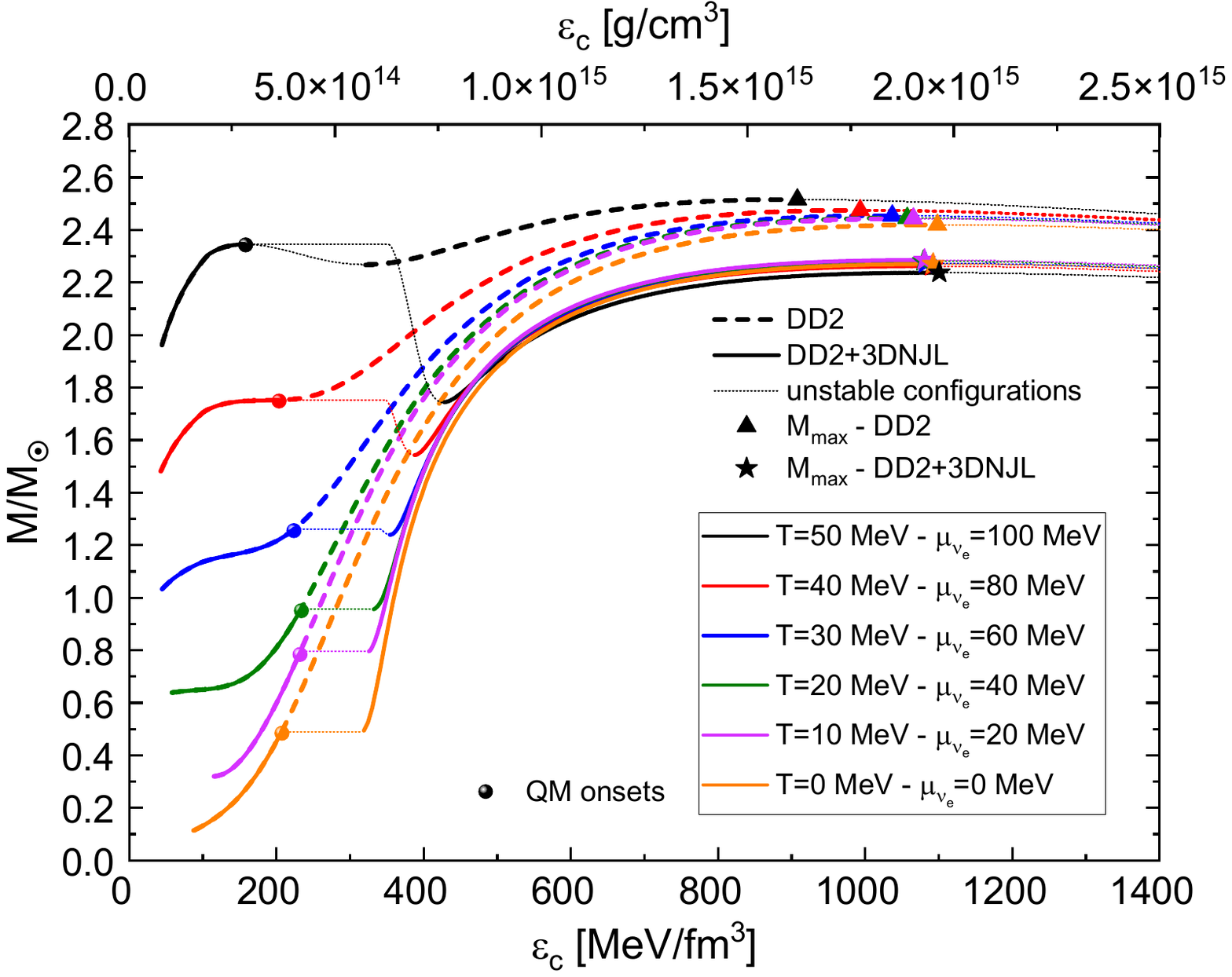}
    \caption{ Mass-central density relations from hybrid EoS for different values of $T$ and $\mu_{\nu_{e}}$. Short-dashed lines correspond to crust+DD2 EoS while solid lines also include QM. For all the cases, dotted lines show the unstable regions. Solid dots display the QM onsets and stars (triangle) symbols indicate the M$_{max}$ locations for hybrid (hadronic) EoS.}
    \label{fig10}
\end{figure}

In Fig.~\ref{fig10}, the mass-central energy density relations of the hadronic (short dashed lines) and hybrid (solid lines) configurations are displayed. The regions of unstable configurations, where the slope of the lines is zero or negative, are represented by thinner short dotted lines for all the cases. The maximum masses are indicated by triangular or star-like symbols for hadronic or hybrid configurations, respectively. Finally, the pure QM onsets are located by solid dot symbols. It is relevant to mention that at $T \geq 30$ MeV, immediately after the phase transition plateau, an unstable region appears which increases with $T$.

\begin{figure}[h]
    \centering    \includegraphics[width=0.48\textwidth]{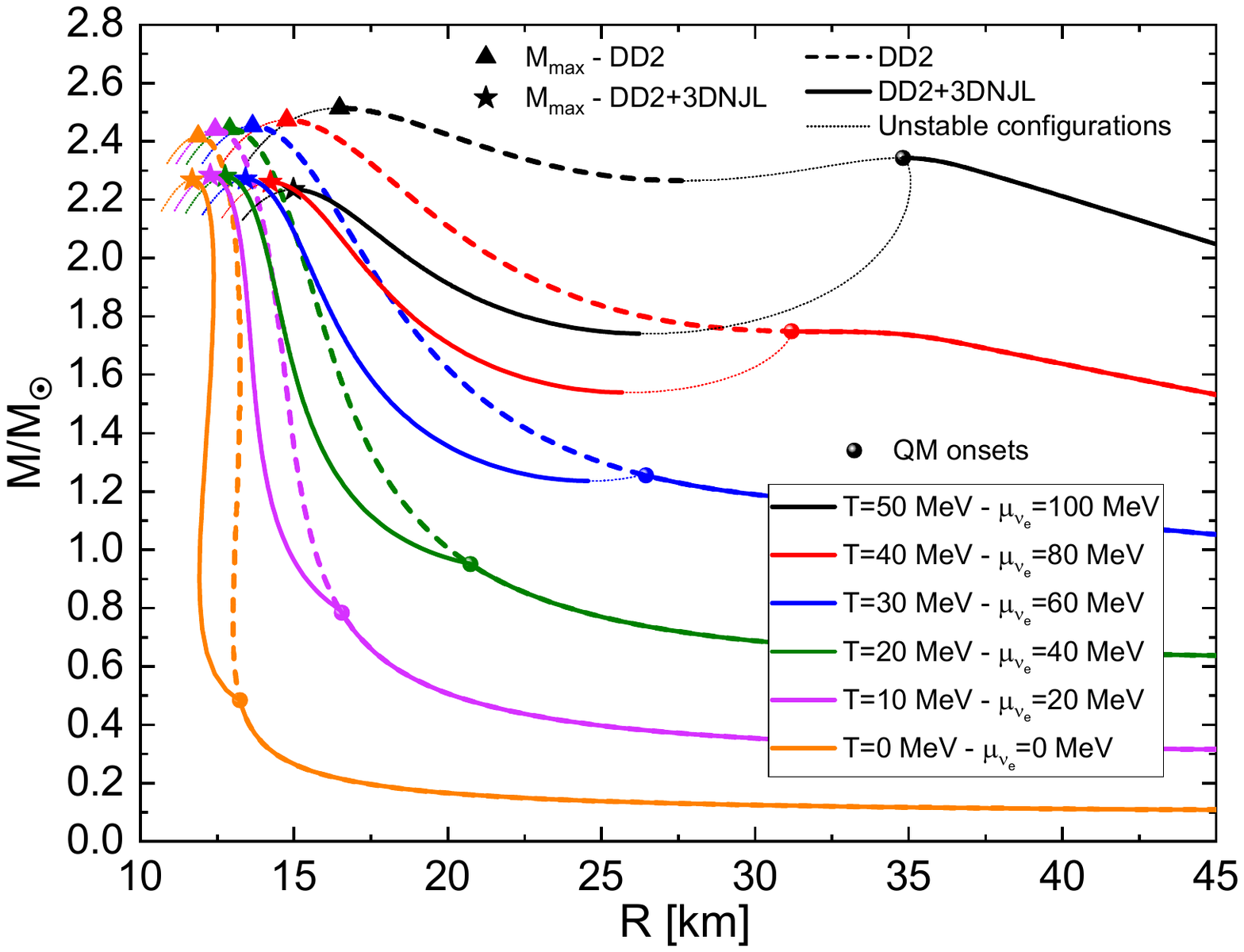}
    \caption{Mass-Radius relations from hybrid EoS for different values of $T$ and $\mu_{\nu_{e}}$. Short-dashed lines correspond to crust+DD2 EoS while solid lines also include QM. For all the cases, dotted lines show the unstable regions. Solid dots display the QM onsets and stars (triangle) symbols indicate the M$_{max}$ locations for hybrid (hadronic) EoS.
    }
    \label{fig11}
\end{figure}

In Fig.~\ref{fig11}, we present the mass-radius relationships for the hybrid compact object configurations (solid lines). For comparison, the corresponding pure hadronic configurations are included (dashed lines). As mentioned earlier, various temperatures have been considered, along with their corresponding $\mu_{\nu_{e}}$ values. The initial configuration for hybrid compact objects, represented by solid dots, shows an increase in both, radius and mass with rising temperature and $\mu_{\nu_{e}}$. We observe that configurations with low radii and maximum mass are located around $2.3 M_{\odot}$ for hybrid stars (star-like symbols) and around $2.5 M_{\odot}$ for pure hadronic stars (triangular symbols). Additionally, it can be seen that the $M_{\rm max}$ locations exhibit a marginal variation for different $T$ and $\mu_{\nu_{e}}$ values. The corresponding radii, however, increase by a few kilometers with $T$ and $\mu_{\nu_{e}}$. It is important to note that at $T \sim 30$ MeV and above, thermal twin configurations appear, consisting of a hybrid compact object and a corresponding hadronic counterpart with the same mass but a significantly larger radius. Similar results have been obtained earlier in Ref. \cite{Hempel:2015vlg} for the class of hybrid EOS where the hadronic phase is described by the STOS EOS \cite{Shen:1998gq, Shen:2011qu} and for the quark matter phase, a bag model has been adopted \cite{Sagert:2008ka, Fischer:2010zzb}. Based on the previously mentioned stability criteria, we can observe (indicated by thinner, short, dotted lines) regions of unstable configurations for $T \geq 30$ MeV in the initial portion of the hybrid star regions. Similarly, for purely hadronic configurations at $T \gtrsim 50$ MeV.

\section{Summary and Conclusions}
\label{SC}

In this study, we have conducted a comprehensive analysis of the low-temperature QCD phase diagram and hybrid protoneutron star configurations including the possibility of neutrino trapping.

The hadronic phase is characterized by the DD2 EOS, incorporating in addition to the density dependence on both temperature and a chemical potential of electron neutrinos trapped within the system.
The quark matter phase is described by a nonlocal quark model with an `instantaneous' form factor, as previously introduced in~\cite{Contrera:2022tqh} at zero temperature. The model's input parameters were carefully selected to align with modern multi-messenger observational data. Initially, we meticulously examined the quark matter phase diagram, both with and without the influence of trapped neutrinos. Our analysis led us to the conclusion that neutrinos have a minimal impact on the phase diagram. 
We assumed a linear relationship between the neutrino chemical potential and temperature, alongside the ansatz $\mu_{\nu_{e}} = \alpha T$, with $\alpha = 2$. Our findings indicate that up to $\alpha = 2.4$ hybrid configurations persist consistently for temperatures up to 50 MeV. Nevertheless, when considering a slightly steeper linear relationship between $T$ and $\mu_{\nu_{e}}$ characterized by a greater $\alpha$ value, the hybrid configurations can not be obtained at this temperature, giving rise to pure hadronic stellar objects. We also observed that the temperature has no significant influence on the maximum mass of the hybrid star sequence. However, the radius of the `hot' compact object increases. An additional effect related to the temperature and neutrinos is observed for temperatures above 30 MeV: stable thermal twin branches emerge, with one component of the equal-mass pair exhibiting a significantly larger radius.
The occurrence of thermal twins in the mass-radius diagram indicates a softening of the quark matter phase which in a dynamical scenario of supernova collapse or neutron star mergers may result in black hole formation, i.e. a ``failed supernova''.

\begin{figure}[h]
    \centering    \includegraphics[width=0.48\textwidth]{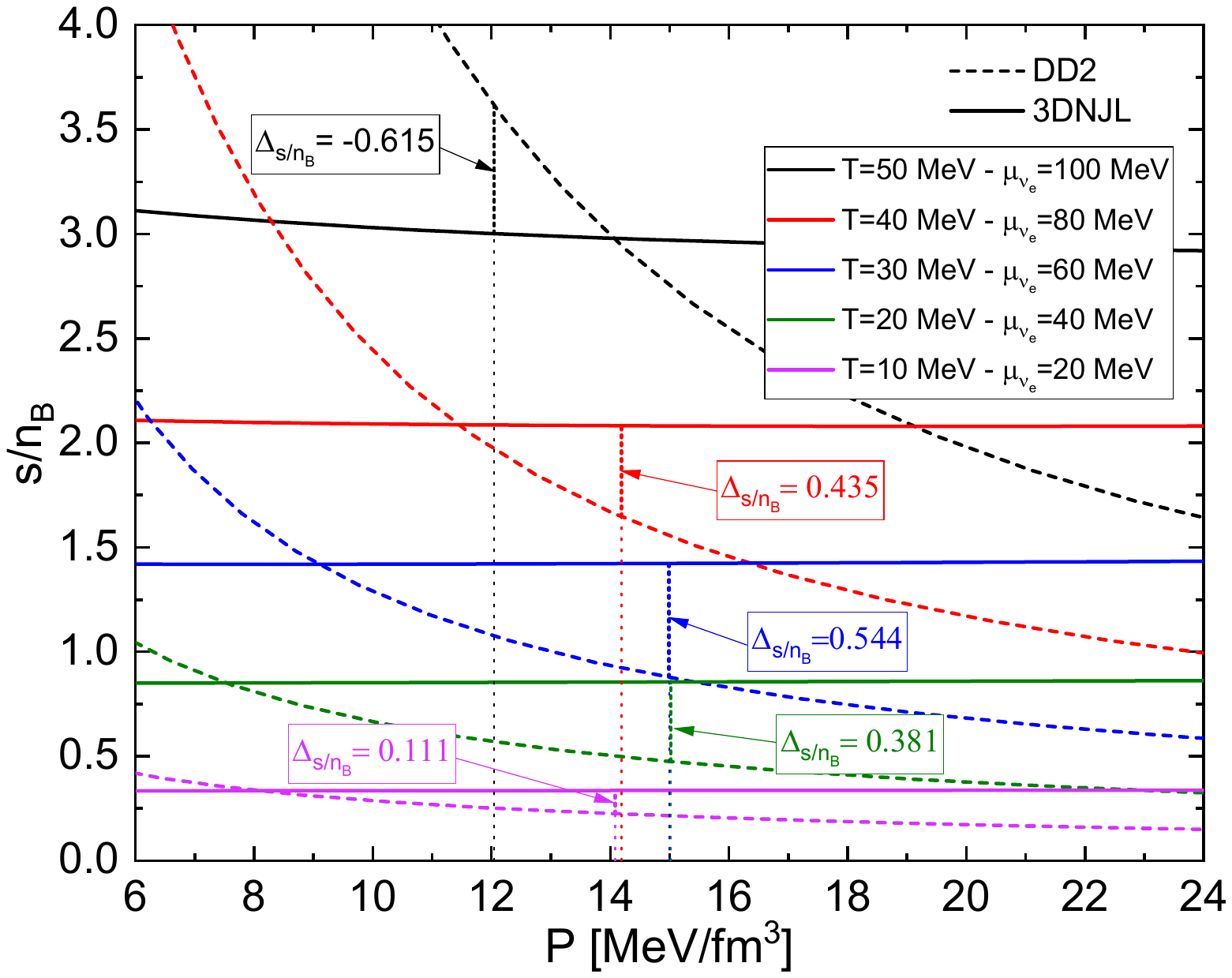}
    \caption{Entropy per baryon vs pressure in the region of critical pressures obtained from Fig.~\ref{fig8} (a), indicated by vertical lines. Dashed lines are for the hadronic (DD2) phase while solid lines are for the QM (3DNJL) phase of the present work.
    }
    \label{fig12}
\end{figure}

Summarizing our investigation, we find that the hybrid EOS model with color superconducting quark matter exhibiting thermal twin stars falls in the same class of hybrid EOS as, e.g., the STOS-B EOS with B-parameters in the range of B139 and B165 \cite{Hempel:2015vlg}, where the entropy per baryon in an isothermal transition decreases. Following the discussion in \cite{Ivanytskyi:2022wln}, this behavior at the deconfinement transition is called {\it enthalpic} transition (see also Fig. 10 of Ref.~\cite{Jakobus:2022ucs}), which was also found for the phase diagram of hybrid EOS \cite{Ivanytskyi:2022wln}
constructed with the confining relativistic density functional model for color superconducting quark matter \cite{Ivanytskyi:2022oxv}.
Such models are likely to produce thermal twins. We want to conjecture that the property of thermal twin configurations is a necessary (but not sufficient) condition for the deconfinement transition at finite temperatures being enthalpic.

On the other hand, as has been discussed in \cite{Fischer:2017lag, Jakobus:2022ucs}, the {\it entropic} transition, which is found for the class of DD2F-SF models without color superconductivity, leads to a rather stiff quark matter core in the hot protoneutron star which entails a strong second shock in a supernova simulation due to the deconfinement transition. It was shown, e.g., in \cite{Fischer:2017lag, Kuroda:2021eiv} this triggers the successful explosion of massive progenitor stars. With this in mind, we have conducted a comparison of the entropy per baryon in the vicinity of the QM onsets shown in Fig.~\ref{fig12} and observed that at $T=50$ MeV, the disparity between the hadronic and QM entropy per baryon, as denoted by $\Delta_{s/n_{B}}=[s/n_{B}]_{QM}-[s/n_{B}]_{HM}$, has a negative value. This corresponds to an enthalpic transition, consistent with the proposed conjecture that the thermal twins appear to be linked to the enthalpic transition (see Ref. \cite{Jakobus:2022ucs} for details). However, at $T=30$ and $40$ MeV, where we observed thermal twin configurations, the phase transition was not enthalpic.

Further detailed investigations should be performed to pin down the possible relationship between the entropic or enthalpic character of the deconfinement transition in the QCD phase diagram and the explodability of supernovae. The thermal twin star property of a hybrid EOS is a facet of this challenging task.

\subsection*{Acknowledgements}
A.G.G., G.A.C., and J.P.C. would like to acknowledge
CONICET, ANPCyT and UNLP (Argentina) for financial support under grants No. PIP 2022-2024 GI - 11220210100150CO, PICT19-00792 and X960, respectively.
D.B. was supported by NCN under grant No. 2019/33/B/ST9/03059.

\begin{appendix}

\section{Details of the nonlocal model for quark matter}
\label{sect:details_QM}

In this Appendix, we show some explicit expressions corresponding to the nonlocal chiral quark model considered in Sec.~\ref{sect:3DFFqm}.

From Eqs. (\ref{gapeq}) the gap equations for the mean fields $\bar{\sigma}$ and $\bar{\Delta}$ together with the constraint equation for the mean-field $\bar\omega$ read
\begin{eqnarray}
\bar\sigma &=& 2 G_s \int \frac{d^3\vec{p}}{(2 \pi)^3}
\frac{g(\vec{p})\ M(\vec{p})}{E}  \\
& &\times \sum_{{c,\kappa=\pm}}
 \frac{ \bar{E}^\kappa_c}{\epsilon^\kappa_c}\left[ 1 - n_F \left(\tfrac{\epsilon^\kappa_c + \delta\tilde\mu_c}{T}\right)
- n_F\left(\tfrac{\epsilon^\kappa_c - \delta\tilde\mu_c}{T}\right) \right] ,\nonumber
\label{gap_sig}
\end{eqnarray}

\begin{eqnarray}
\bar\Delta &=&  2 G_D  \int \frac{d^3\vec{p}}{(2 \pi)^3}
g^2(\vec{p}) \ \bar\Delta   \\
&&\times  \sum_{\kappa=\pm}
\left\{ \frac{2}{\epsilon^\kappa_r}\left[ 1 - n_F\left(\tfrac{\epsilon^\kappa_r + \delta\tilde\mu_r}{T}\right)
- n_F\left(\tfrac{\epsilon^\kappa_r - \delta\tilde\mu_r}{T}\right) \right] \right\},\nonumber
\label{gap_del}
\end{eqnarray}
and
\begin{eqnarray}
\bar\omega &=& 2 G_V \int \frac{d^3\vec{p}}{(2 \pi)^3}  \\
& &\times \sum_{{c,\kappa=\pm}}
 \kappa \frac{ \bar{E}^\kappa_c}{\epsilon^\kappa_c}\left[ 1 - n_F\left(\tfrac{\epsilon^\kappa_c + \delta\tilde\mu_c}{T}\right)
- n_F\left(\tfrac{\epsilon^\kappa_c - \delta\tilde\mu_c}{T}\right) \right] ,\nonumber
\label{gap_ovec}
\end{eqnarray}
where $n_F(x)=(1+\exp(x))^{-1}$ is the Fermi distribution function.
The solutions for Eqs. (\ref{gap_sig})-(\ref{gap_ovec}) in the vacuum, at $T=\mu=0$, are denoted with the subscript $0$ as $\bar \sigma_0$, $\bar \Delta_0$ and $\bar \omega_0$, respectively.
In the vacuum $\bar \Delta_0$ and $\bar \omega_0$ are zero, so the vacuum thermodynamic potential reads
\begin{eqnarray}
\Omega^{MFA}_0 = \frac{ \bar
\sigma^2_0 }{2 G_S} - 12 \int \frac{d^3 \vec{p}}{(2\pi)^3} \; E_0/2.
\label{mfavac}
\end{eqnarray}
Notice that in the above expression $E_0^2 = \vec{p}~^2 + M_0^2(\vec{p})$ where $M_0(\vec{p}) = m_c + g(\vec{p}) \bar\sigma_0$.
The integral in Eq.~(\ref{mfaqmtp}) turns out to be ultraviolet divergent because of the zero-point energy terms. Since this is exactly the divergence of Eq.~(\ref{mfavac}), a successful regularization scheme consists just in the vacuum subtraction
\begin{equation}
\Omega^{MFA}_{reg} = \Omega^{MFA} - \Omega^{MFA}_0 \,.
\label{eq:omreg}
\end{equation}
Finally, using Eq. (\ref{densities}) the quarks and electron number densities, $n_{fc}$ and $n_e$ respectively, are given by
\begin{eqnarray}
n_{fc} &=&
 - \int \frac{d^3 \vec{p}}{(2\pi)^3} \;  \\
&\times& \sum_{\kappa=\pm} \left\{\left[  n_F\left(\tfrac{\epsilon_c^\kappa + \delta \tilde\mu_c}{T}\right) -  n_F\left(\tfrac{\epsilon_c^\kappa - \delta \tilde\mu_c}{T}\right)\right] (\delta_{uf} - \delta_{df}) \right.\nonumber \\
&-& \left. \kappa \;
\frac{\bar{E}_c^\kappa}{\epsilon_c^\kappa} \left[ 1 -  n_F\left(\tfrac{\epsilon_c^\kappa +  \delta\tilde\mu_c}{T}\right) - n_F\left(\tfrac{\epsilon_c^\kappa -  \delta\tilde\mu_c}{T}\right)
 \right]\right\}, \nonumber
\label{rhofr}
\end{eqnarray}
and
\begin{eqnarray}
n_e &=& - 2  \;  \sum
\int \frac{d^3 \vec{p}}{(2\pi)^3} \;  \\
& &\times \left[ n_F\left(\frac{\epsilon_e +  \; \mu_e}{T}\right) - n_F\left(\frac{\epsilon_e -  \; \mu_e}{T}\right)\right]. \nonumber
\label{rholep}
\end{eqnarray}
Note that the fermion density has only the contribution of the first term of the regularized thermodynamic potential (\ref{eq:omreg}), since the second one has no dependence on the chemical potentials.

Let us explicitly show the entropy density $s = s_q + s_{lep}$ contribution both for quarks
\begin{eqnarray}
s_q &=&  2 \sum_{c,\kappa,\lambda=\pm} \int  \frac{d^3 p}{(2\pi)^3} \left\{
\ln \left[ 1 \; + \; e^{- \frac{\epsilon _c^\kappa + \lambda \delta\tilde\mu_c}{T}}\right] \right. \nonumber\\
&+& \left.
\left( \frac{\epsilon _c^\kappa + \lambda \delta\tilde\mu_c}{T} \right)
n_F\left(\tfrac{\epsilon _c^\kappa + \lambda \delta\tilde\mu_c}{T}\right) \right\} \ ,
\label{entropy}
\end{eqnarray}
(remember that for $c=b$: $\Delta=0$ and $\epsilon_b^\pm=\bar{E}_b^\pm$) and leptons
\begin{eqnarray}
s_{lep} &=& 2  \;  \sum_{\lambda =\pm}
\int \frac{d^3p}{(2 \pi)^3} \nonumber \\
&\times& \left\{ \ln \left[
1 \; + \; e^{- \tfrac{\epsilon_e + \lambda \; \mu_e}{T}}\right]
+ \; \frac{\epsilon_e + \lambda \; \mu_e }{T}\ n_F\left(\tfrac{\epsilon_e + \lambda \; \mu_e }{T}\right) \right\} \nonumber \\
&+&
\left(\frac{\mu_{\nu_e}^2\ T}{6} + \frac{7\pi^2\ T^3}{90} \right) \ .
\label{lep}
\end{eqnarray}

\end{appendix}

\bibliography{3DFF_finiteT}

\end{document}